\documentclass[letterpaper, 10 pt, journal, twoside]{IEEEtran}
\usepackage{microtype}
\usepackage{graphicx}
\usepackage{subfigure}
\usepackage{booktabs}
\usepackage{epsfig}
\usepackage[labelfont=bf]{caption}
\usepackage{indentfirst}
\usepackage{amsmath}
\usepackage{amssymb}
\usepackage{mathtools}
\usepackage{footmisc}
\usepackage{xcolor}
\usepackage[pagebackref,breaklinks,colorlinks]{hyperref}
\definecolor{maroon}{rgb}{0.8, 0.25, 0.33}

\usepackage[capitalize,noabbrev]{cleveref}
\usepackage{float}

\begin{document}

\title{\LARGE \bf
MA-DV$^2$F: A Multi-Agent Navigation Framework using Dynamic Velocity Vector Field}

\author{Yining Ma$^{1,2}$, Qadeer Khan$^{1,2}$ and Daniel Cremers$^{1,2}$%
\thanks{$^{1}$Chair of Computer Vision \& Artificial Intelligence, School of Computation, Information and Technology, Technical University of Munich. \\ Contact: {\tt\footnotesize \{yining.ma, qadeer.khan, cremers\}@tum.de}}%
\thanks{$^{2}$Munich Center for Machine Learning (MCML).}%
}

\maketitle

\begin{abstract}
In this paper we propose MA-DV$^2$F: Multi-Agent Dynamic Velocity Vector Field. It is a framework for simultaneously controlling a group of vehicles in challenging environments. DV$^2$F is generated for each vehicle independently and provides a map of reference orientation and speed that a vehicle must attain at any point on the navigation grid such that it safely reaches its target. The field is dynamically updated depending on the speed and proximity of the ego-vehicle to other agents. This dynamic adaptation of the velocity vector field allows prevention of imminent collisions.  Experimental results show that MA-DV$^2$F outperforms concurrent methods in terms of safety, computational efficiency and accuracy in reaching the target when scaling to a large number of vehicles. Project page for this work can be found here: \href{https://yininghase.github.io/MA-DV2F/}{https://yininghase.github.io/MA-DV2F/}
\end{abstract}

\section{INTRODUCTION} \label{sec:introduction}
The task of multi-agent navigation has attracted widespread attention in recent years due to myriad applications in areas such as search and rescue missions \cite{10517396}, area exploration \cite{10184313}, pickup and delivery services \cite{10207838}, warehouses \cite{10113719}, self-driving cars \cite{multivehicle2023} etc. The task of multi-agent path finding/navigation involves simultaneously directing a group of vehicles from their initial position to their desired destination  while avoiding collisions with other agents. The task is known to be NP-hard even in the discrete setting \cite{nebel2020computational}. An ideal algorithm must find the optimal solution in limited time. This leads to contradictory goals, since determining the optimal solution requires searching a larger solution space, which necessitates more time. In structured environments such as indoor spaces, prior knowledge and understating of the layout impose constraints, that can reduce the solution search space. In unstructured environments, there are no such constraints. This allows an algorithm the flexibility to find a solution. However, since the search space is much larger, there is no guarantee that the solution found is optimal. The problem is further exacerbated when the search space is continuous and agents are non-holomonic  vehicles. The constraints arising from the vehicle kinematics add to the complexity. 

There have been various techniques and heuristics attempting to find (near-)optimal trajectories for multiple agents. The methods can be divided into two primary categories: 1) Learning based data driven methods \cite{zhang2023gcbf,zhang2024gcbf+} and 2) Search/optimization based methods \cite{wen2022cl,10628993}. Learning based algorithms involve training a neural network on data, with the understanding that the network will generalize at inference time. The training data should encompass all the situations that the model is expected to encounter at test time. This necessitates a large amount of training samples. The greatest challenge with large training samples is determining the corresponding supervised labels for training; that might be too tedious to obtain. In contrast to supervised learning, an alternate would be to train using reinforcement learning (RL)  \cite{sutton2018reinforcement}, where the model explores the environment and acquires rewards or penalties depending on the actions taken. The model then exploits this experience to execute the correct control actions at test time. However, RL algorithms tend to be more sample inefficient than supervised methods. In contrast, optimization \cite{multiagent2023} or search based \cite{sharon2015conflict} methods involves simultaneously optimizing trajectories for multiple vehicles. As the number of vehicles are added, the complexity of the optimization/search becomes intractable making it infeasible for scaling to a large number of vehicles \cite{li2020graph,ma2024enhancing}.

In this paper, we propose Multi-Agent Dynamic Velocity Vector Field (MA-DV$^2$F), which generates vectors for the orientation and reference speed for every vehicle on the map. The vehicles then just need to follow in the direction of their respective velocity vector fields to successfully reach their destinations.  The vector field for each vehicle is generated independently and can be adapted dynamically depending on the vehicle's proximity to other agents (neighbouring vehicles or obstacles).  Decoupling reduces the complexity of adding vehicles \& allows for parallel generation of DV$^2$F of each vehicle, thereby increasing the throughput. An added benefit of our approach is that the generated DV$^2$F can be used to train a learning based graphical neural network (GNN) in a self-supervised manner. This self-supervised learning based approach neither requires tedious labeling of data nor necessitates sample inefficient environment exploration.  We test our framework under challenging collision prone environments.  A scenario is regarded to be challenging if trajectories of different vehicles considered independently from other agents intersect at multiple places at the same time. This would necessitate a  collision avoidance maneuver for safe navigation towards the target.   

\begin{figure}[!t]
\centering
\includegraphics[width=0.98\linewidth]{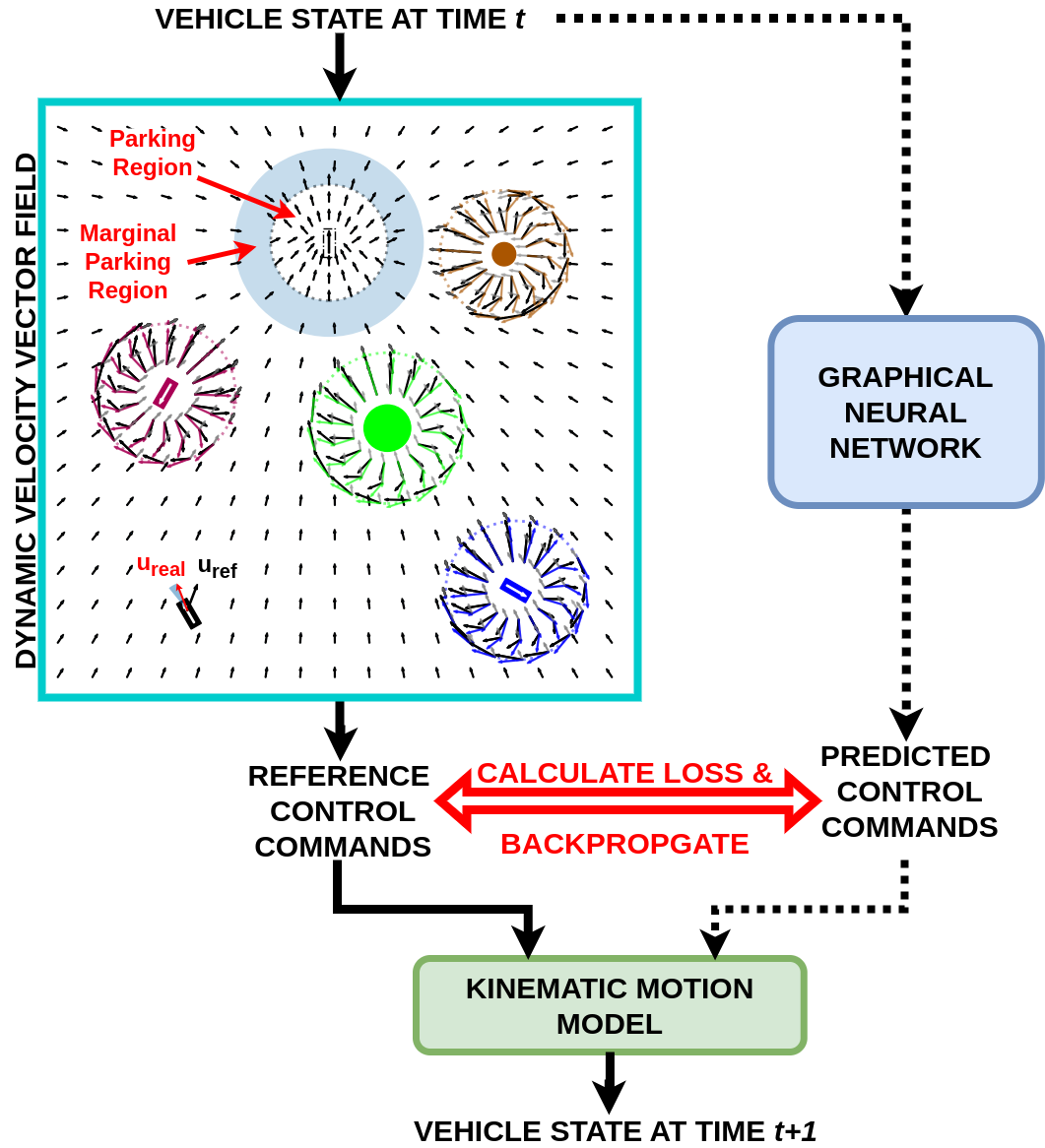}
\caption{Shows the pipeline for MA-DV$^2$F. The state of all vehicles at time $t$ is used to create the dynamic velocity vector field (DV$^2$F) from which the reference control commands are determined. These commands are in turn used to determine the next state at time $t+1$ using the kinematic motion model. Note that the reference control commands can optionally be used to train a GNN in a self-supervised manner (indicated by dotted arrows). Meanwhile, the DV$^2$F is shown for the black ego-vehicle with 2 other vehicles (\textcolor{blue}{\textbf{blue}} \& \textcolor{maroon}{\textbf{maroon}}) and 2 obstacles (\textcolor{green}{\textbf{green}} \& \textcolor{brown}{\textbf{brown}}) in the scene. The target state for the black ego-vehicle is shown by the dotted rectangle with a solid arrow at the top of the map. The dotted circle around the target state is the parking region. The black arrows indicate the ideal reference orientation as unit vectors which the ego-vehicle should attain at each different position on the map. The colored dotted circles around other vehicles and obstacles show the collision avoiding regions for the black ego-vehicle. Each black arrow in these regions is composed of an attractive target approaching component (gray arrow) and a repulsive collision avoidance component (colored arrow). Note that due to kinematic constraints, the reference orientation ($\mathbf{u}_\mathbf{ref}$) might not be attainable for the ego-vehicle at its existing location. The shaded wedge in front of the ego-vehicle shows the region of reachable orientation at the next time step. The real orientation (\textcolor{red}{$\mathbf{u}_\mathbf{real}$}) is therefore the physically attainable orientation by the ego-vehicle that is closest to the reference. An example of the dynamic velocity vector field is shown in the project page: \href{https://yininghase.github.io/MA-DV2F/\#VD}{https://yininghase.github.io/MA-DV2F/\#VD}. Note that DV$^2$F for the neighboring \textcolor{blue}{\textbf{blue}} \& \textcolor{maroon}{\textbf{maroon}} vehicles is likewise created separately (not shown in this figure).}
\label{fig:pipeline overview}
\end{figure}

Fig. \ref{fig:pipeline overview} shows the pipeline for both the MA-DV$^2$F (left branch, solid arrows) and optional training of the the self-supervised GNN counterpart (right branch, dotted arrows). The input is a continuous state representation of all the multiple vehicles and the outputs are the corresponding continuous control variables. The vehicles are non-holonomic with rotation radius determined by the kinematic model. 

We summarize the contribution of this letter as follows:
\begin{itemize}
  \item Our proposed MA-DV$^2$F outperforms other concurrent learning and search based approaches for the task of multi agent navigation in challenging, collision prone environments.
  \item Even the self-supervised learning based counterpart of MA-DV$^2$F scales better than other learning and search based methods. 
  \item MA-DV$^2$F can determine the solutions orders of magnitude faster than other SOTA search based approaches.
  \item We release the complete code of MA-DV$^2$F on the project page here: \href{https://yininghase.github.io/MA-DV2F/}{https://yininghase.github.io/MA-DV2F/}. 
\end{itemize}
The project page also contains additional supplementary information such videos better depicting the operation of our method in comparison with other approaches, details of the challenging scenarios, dynamics of the velocity field at different regions on the navigation grid etc.
\section{Related Work}\label{sec:related_work}

\cite{khatib1985real} proposed using Artificial Potential Fields (APF) for trajectory planning.  An agent is subjected to an attractive potential force towards the target which serves as the sink and a repulsive potential force away from obstacles \cite{reda2024path,loganathan2023systematic}.  However, a common problem with such methods is their propensity to get stuck in local minima \cite{madridano2021trajectory,hou2023large} when the attractive force from the target is cancelled out by the repulsive force arising from an another agent for e.g. when the ego-vehicle is symmetrically aligned with other agents to the target \cite{iswanto2019artificial} and thus leading to a bottleneck situation. We break such bottlenecks by enforcing the vehicles to move in the clockwise direction.

\cite{sharon2015conflict} proposed a two level tree based search algorithm for multi-agent path finding. However, the tree may grow exponentially, making the search inefficient. This is because multi-agent path planning methods on trees and graphs are known to be NP-hard \cite{yu2013structure} since the search space grows exponentially as the number of agents rise \cite{sharon2015conflict}. Nevertheless, \cite{li2020graph} used \cite{sharon2015conflict} to generate expert data for training a GNN model that can scale up to more vehicles than trained on. \cite{wai2018multi}  uses RL and proposes a dense reward function to encourage environmental exploration. However, the need for exploration tends to make the learning sample-inefficient \cite{8686348,electronics9091363} particularly when compared with imitation learning approaches \cite{pmlr-v78-dosovitskiy17a}. \cite{8661608} rather combines RL for single agent path planning with imitation learning to learn actions that can influence other agents. All approaches described  above work either on a discrete grid, discrete action space, assume holomonic robots or their combination. 

CL-MAPF \cite{wen2022cl} uses a Body Conflict Tree to describe agent collision scenarios as spatiotemporal constraints. It then applies a Hybrid-State A* Search algorithm to generate paths satisfying both kinematic and spatiotemporal constraints of the vehicles. However, under challenging test scenarios with vehicles crowding together, the algorithm takes long to search for a solution and can easily time out. To find feasible solution for large-scale multi-vehicle trajectory planning, CSDO \cite{10628993} first searches for a coarse initial guess using a large search step. Then, the Decentralized Quadratic Programming is implemented to refine this guess for minor collisions. GCBF+ \cite{zhang2024gcbf+} based on GCBF \cite{zhang2023gcbf} aims to provide safety guarantees utilizing  control barrier functions (CBF). A Graphical Neural Network is trained to learn agent control policy.

\section{Framework}\label{sec:framework}

\subsection{Problem Description}\label{subsec:prob_def}

We aim to solve the task of multi-agent navigation in unconstrained environments. Given $N_{veh}$ dynamic vehicles and $N_{obs}$  static obstacles in the scene, the task is to ensure each vehicle reaches its desired destination while avoiding collision with other agents. The state vector for ego-vehicle $i$ ($1 \le i \le N_{veh}$)  at current time $t$ can be represented as $\mathbf{s}_{t}^{({i})} =  [x_{t}, y_{t}, \theta_{t}, v_{t}, x_{tar}, y_{tar}, \theta_{tar}]^{T}$, where $x_{t}$ and $y_{t}$ shows the position, $\theta_{t}$, the orientation and $v_{t}$ the speed at current time $t$. Meanwhile, $x_{tar}$ and $y_{tar}$ are coordinates of the target position, and $\theta_{tar}$ is the target orientation. Each ego-vehicle is controlled by a control vector $\mathbf{c}_{t}^{({i}_{veh})} =  [p_{t}, \varphi_{t}]^{T}$, where $p_{t} \in [-P, P]$ and $\varphi_{t} \in [-\Phi, \Phi]$ are the pedal acceleration and steering angle, limited in magnitude by $P$ and $\Phi$ respectively. The obstacle $k$ ($1 \le k \le N_{obs}$) is represented by a state vector $\mathbf{s}^{k} = [x, y, r]^{T}$, where $x$ and $y$ is the position, and $r$ is the radius of the circle circumscribing all vertices of the obstacle. The kinematics of the vehicles is modeled using the bicycle model \cite{wang2001trajectory}. 
\begin{equation}
\begin{split}
x_{t+1} & = x_{t} + v_{t} \cdot \cos(\theta_{t}) \cdot \Delta t \\ y_{t+1} & = y_{t} + v_{t} \cdot \sin(\theta_{t}) \cdot \Delta t \\ \theta_{t+1} & = \theta_{t} + v_{t} \cdot \tan(\varphi_{t}) \cdot \gamma \cdot \Delta t \\
v_{t+1} & = \beta \cdot v_{t} +  p_{t} \cdot \Delta t 
\end{split}
\label{equ:vehicle kinematic}
\end{equation}

It describes how the equations of motion can be updated in time increments of $\Delta t$ assuming no slip condition, valid under low or moderate vehicle speed $v$ when making turns. Note that $\beta$ and $\gamma$ are the tuneable hyper-parameters modeling the environment friction and vehicle wheelbase respectively.

We create a velocity vector field for each vehicle, which in turn can dynamically generate reference control variables. Generation of the velocity field can be divided into two stages:
\begin{itemize}
\item Estimation of reference orientation of  ego-vehicle for every position in the map (See Subsection \ref{subsec:orientation}).
\item Estimation of reference speed for the corresponding positions. (See Subsection \ref{subsec:speed}).
\end{itemize}

The orientation which a vehicle should possess at any position on the map is referred to as the \textit{Reference Orientation} and is represented as a unit vector. The black arrows in Fig. \ref{fig:pipeline overview} show the reference orientation for the black ego vehicle at different positions in the map. It can be seen that the arrows attract the ego-vehicle towards its target while repelling it away from the other vehicles and obstacles. Note that the current orientation of the ego-vehicle is not aligned with the reference orientation at its existing position. Therefore, we ought to find the control variables which align the ego-vehicle with the reference orientation. Apart from the reference orientation at each position in the map, the ego-vehicle also has a reference speed which should be attained by the control variables. Lastly, note that apart from the black vehicle, a separate reference orientation map is also created for the blue and maroon vehicles present in Fig. \ref{fig:pipeline overview} (not shown in the figure).

Hence, our task of multi-agent navigation simplifies  to finding the reference orientation maps and corresponding reference speed for each vehicle independently. In the subsequent sub-sections, we describe how the reference orientation and speed are estimated.

\subsection{Estimation of Reference Orientation}\label{subsec:orientation}

We first define some frequently used functions and vectors before we begin discussion of reference orientation estimation. The vector function $\mathbf{f_{uni}}(\mathbf{a})$ is the function which takes a non-zero vector $\mathbf{a}$ as input and divides by its magnitude to convert it into a unit vector.

Other scalar functions include $f_{sgn}(a)$ and $f_{pos}(a)$ which both output 1 if the scalar input $a$ is positive. However, $f_{sgn}(a)$ outputs -1, while $f_{pos}(a)$ outputs 0 when $a$ is negative. We now define the vector from the next position $(x^{(i)}_{t+1}, y^{(i)}_{t+1})$ of ego vehicle $i$ to

\begin{itemize}
 \item its target position $(x^{(i)}_{tar}, y^{(i)}_{tar})$ as $\mathbf{X}^{(i)}_{tar} = [x^{(i)}_{tar}-x^{(i)}_{t+1}, y^{(i)}_{tar}-y^{(i)}_{t+1}]^{T}$
 \item the position $(x^{(k)}_{obs}, y^{(k)}_{obs})$ of the static obstacle $k$ as $\mathbf{X}^{(i)}_{obs_{k}} = [x^{(k)}_{obs}-x^{(i)}_{t+1}, y^{(k)}_{obs}-y^{(i)}_{t+1}]^{T}$
 \item next position $(x^{(j)}_{t+1}, y^{(j)}_{t+1})$ of another vehicle $j$ as $\mathbf{X}^{(i)}_{veh_{j}} = [x^{(j)}_{t+1}-x^{(i)}_{t+1},y^{(j)}_{t+1}-y^{(i)}_{t+1}]^{T}$
\end{itemize}

The  unit vector along the Z-axis is $\mathbf{Z}=[0, 0, 1]^{T}$, while the unit orientation vector of the ego vehicle at the current target  states are given by $\mathbf{U}^{(i)}_{t} = [\cos(\theta^{(i)}_{t}) , \sin(\theta^{(i)}_{t}) ]^{T}$ and $\mathbf{U}^{(i)}_{tar} = [\cos(\theta^{(i)}_{tar}) , \sin(\theta^{(i)}_{tar}) ]^{T}$ respectively.

For the ego vehicle $i$, the target reaching component of the  reference orientation $\mathbf{u}^{(i)}_{tar}$ is defined as:
\begin{equation}
\begin{split}
    \mathbf{u}^{(i)}_{tar} & =
    \begin{cases}
      \mathbf{f_{uni}}(\mathbf{X}^{(i)}_{tar}) \cdot \xi_{tar}^{(i)} & \|\mathbf{X}^{(i)}_{tar}\|_2 > r_{p} \\
     \mathbf{f_{uni}}(\mathbf{U}^{(i)}_{tar} +  \lambda^{(i)}_{tar}  \cdot  \mathbf{f_{uni}}(\mathbf{X}^{(i)}_{tar})) & \text{otherwise}
    \end{cases} \\ 
  \lambda^{(i)}_{tar} & = (\frac{\|\mathbf{X}^{(i)}_{tar}\|_2}{r_{p}} + f_{pos}(\|\mathbf{X}^{(i)}_{tar}\|_2-\epsilon_{p})) \cdot f_{sgn}({\mathbf{X}_{tar}^{(i)}}^{T} \cdot \mathbf{U}^{(i)}_{tar})\\
  \xi_{tar}^{(i)} & = 
  \begin{cases}
    1 & \|\mathbf{X}^{(i)}_{tar}\|_2 \ge 0.5 \cdot v^{2}_{d} + r_{p} \\
    f_{sgn}({\mathbf{X}_{tar}^{(i)}}^{T} \cdot \mathbf{U}^{(i)}_{t}) & \text{otherwise}\\
  \end{cases}\\
\end{split}
\label{equ:reference orientation target approaching}
\end{equation}

where $r_{p}$ is the parking threshold (black dotted circle in Fig. \ref{fig:pipeline overview}) and $0.5 \cdot v^{2}_{d}$ is the marginal parking threshold (shaded blue region in Fig. \ref{fig:pipeline overview}).  $v_{d}$ is the default reference speed, estimation of which is explained in Subsection \ref{subsec:speed}. $\epsilon_{p}$ is the threshold above which an additional term is introduced when the vehicle is within the parking threshold.

Equation \ref{equ:reference orientation target approaching} shows that when the ego-vehicle $i$ is far away from the parking threshold, the reference orientation is in the direction of $\mathbf{X}^{(i)}_{tar}$. However, as the ego-vehicle approaches the target and the velocity is high, then the vehicle might overshoot the target and enter the shaded marginal parking threshold region. In this case the direction of the orientation is flipped using $f_{sgn}({\mathbf{X}_{tar}^{(i)}}^{T}. \mathbf{U}^{(i)}_{t})$. This is to prevent the vehicle from moving in circles. A detailed explanation of this is given in the supplementary file on the project page: \href{https://yininghase.github.io/MA-DV2F/supplementary.pdf}{https://yininghase.github.io/MA-DV2F/supplementary.pdf}.

Equation \ref{equ:reference orientation target approaching} also shows for the condition when the distance $\|\mathbf{X}^{(i)}_{tar}\|_2$ falls below the parking threshold $r_{p}$, and is closer to the target. The ego vehicle should be guided to not only reach the position of the target ($\mathbf{X}^{(i)}_{tar}=0$) but also be aligned with the target orientation ($\mathbf{U}^{(i)}_{tar} = \mathbf{U}^{(i)}_{t}$). The balancing act between these two conditions is handled by the $\lambda^{(i)}_{tar}$ term in Equation \ref{equ:reference orientation target approaching}. Like previously, $f_{sgn}({\mathbf{X}_{tar}^{(i)}}^{T} \cdot \mathbf{U}^{(i)}_{tar})$ in $\lambda^{(i)}_{tar}$ makes sure the reference and target orientations are in the same directions when parking. The $f_{pos}(\|\mathbf{X}^{(i)}_{tar}\|_2-\epsilon_{p})$ term in $\lambda^{(i)}_{tar}$ is designed to expedite the parking behavior when the vehicle position is exactly on the left or the right side of the target and the vehicle orientation is parallel to the target orientation. 

\begin{figure}[ht]
\centering
    \includegraphics[width=\linewidth]{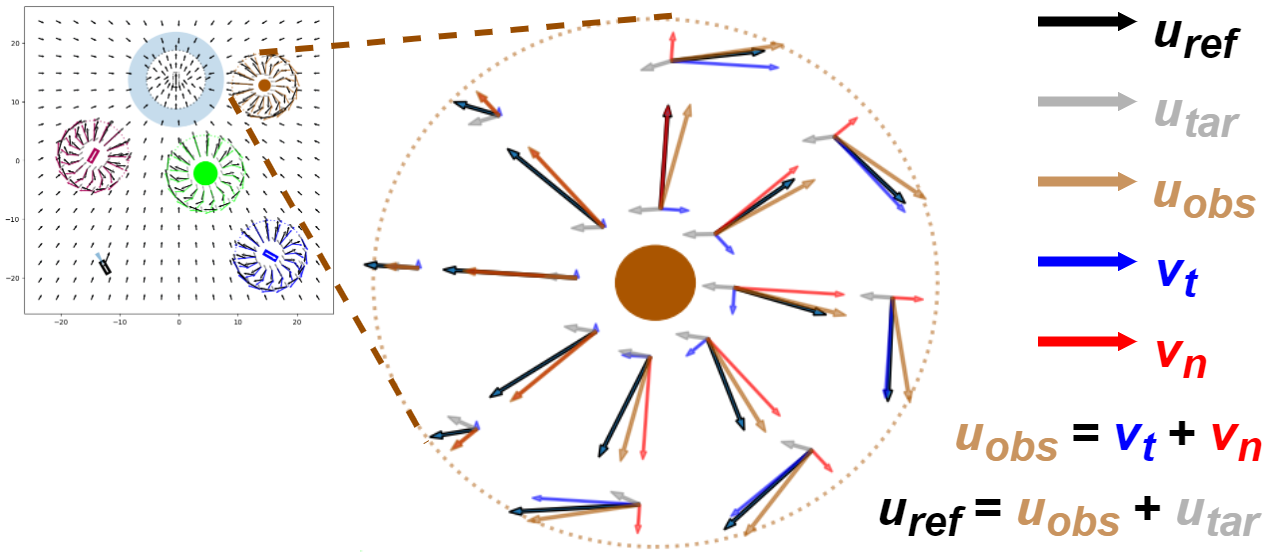}
    \caption{Shows the constituents of the reference orientation vector $\mathbf{u}_{ref}$ near an obstacle. It comprises of a target approaching ($\mathbf{u}_{tar}$) and the collision avoiding ($\mathbf{u}_{obs_{k}}$) components. $\mathbf{u}_{obs_{k}}$ includes $\mathbf{v}_{n}=\mathbf{f_{uni}}(\mathbf{X}^{(i)}_{obs_{k}}) \cdot \alpha^{(i)}_{obs_{k}}$ guiding the vehicle to drive away from the obstacle and $\mathbf{v}_{t}=\mathbf{R}^{(i)}_{obs_{k}} \cdot \beta^{(i)}_{obs_{k}}$ leading the vehicle to go around the obstacle.  When the ego-vehicle is between the obstacle and  target, it is not necessary for it to go around the obstacle and thus $\mathbf{v}_{t} = \mathbf{0}$. The formulation to calculate $\mathbf{u}_{obs_{k}}$ is described in Equation \ref{equ:reference orientation obstacle avoiding}.} 
    \label{fig:reference orientation near an obstacle}
\end{figure}

Besides reaching the target, the ego vehicle $i$ should also avoid collision on its way to the target. This is realized by the collision avoiding component $\mathbf{u}^{(i)}_{coll}$ which comprises of collision avoidance between the ego vehicle $i$ and either static obstacle $k$ ($\mathbf{u}^{(i)}_{obs_{k}}$) or with another vehicle $j$ ($\mathbf{u}^{(i)}_{veh_{j}}$).\\ The equation for determining ($\mathbf{u}^{(i)}_{obs_{k}}$)  is given by:
\begin{equation}
\begin{split}
\mathbf{u}^{(i)}_{obs_{k}} & =
\begin{cases}
\mathbf{f_{uni}}(\mathbf{X}^{(i)}_{obs_{k}}) \cdot \alpha^{(i)}_{obs_{k}} + \mathbf{R}^{(i)}_{obs_{k}} \cdot \beta^{(i)}_{obs_{k}} & \alpha^{(i)}_{obs_{k}} \le 0 \\
\mathbf{0} & \text{otherwise} \\
\end{cases} \\
\alpha^{(i)}_{obs_{k}} & = \|\mathbf{X}^{(i)}_{obs_{k}}\|_2-r_{obs}^{(k)}-r_{veh}-(r_{c} + |v^{(i)}_{t}|) \\
\beta^{(i)}_{obs_{k}} & = f_{pos}({\mathbf{X}^{(i)}_{tar}}^{T} \cdot \mathbf{X}_{obs_{k}}^{(i)}) \cdot (\|\mathbf{X}^{(i)}_{obs_{k}}\|_2-r_{obs}^{(k)}) \\
\mathbf{R}^{(i)}_{obs_{k}} & = \mathbf{f_{uni}}(\mathbf{Z} \times \mathbf{X}^{(i)}_{obs_{k}}) \\
\end{split}
\label{equ:reference orientation obstacle avoiding}
\end{equation}

where $r^{(k)}_{obs}$ is the radius of the static obstacle $k$, $r_{veh}$ is the radius of the smallest circle enclosing the ego-vehicles, $r_{c}$ is the static component, while $|v^{(i)}_{t}|$ is the speed-based dynamic safety margin for collision avoidance between ego vehicle $i$ and obstacle $k$. Higher the vehicle speed $|v^{(i)}_{t}|$, larger is this margin $r_{c}+|v^{(i)}_{t}|$. When the distance between the ego vehicle $i$ and static obstacle $k$ ($\|\mathbf{X}^{(i)}_{obs_{k}}\|_2-r_{obs}^{(k)}-r_{veh}$) is below the collision avoidance margin $r_{c} + |v^{(i)}_{t}|$, then $\alpha^{(i)}_{obs_{k}} \le 0$, and the reference orientation is modified to prevent collision with the static obstacle. Under this condition, the first term $\mathbf{f_{uni}}(\mathbf{X}^{(i)}_{obs_{k}}) \cdot \alpha^{(i)}_{obs_{k}}$ will guide the vehicle to drive away from the static obstacle. However, driving away is not enough as this might cause a bottleneck in cases where the obstacle is symmetrically collinear between the ego-vehicle and its target. We would additionally like the ego-vehicle to drive around the obstacle to reach the target for which the term $\mathbf{R}^{(i)}_{obs_{k}} \cdot \beta^{(i)}_{obs_{k}}$ is relevant. If the ego-vehicle is between the obstacle and target then $\beta^{(i)}_{obs_{k}} = 0$ (since $f_{pos}({\mathbf{X}^{(i)}_{tar}}^{T} \cdot \mathbf{X}_{obs_{k}}^{(i)}) = 0$) and hence  there is no need for the vehicle to drive around the obstacle. However, if that is not the case, then an additional component is added whose direction is given by $\mathbf{R}^{(i)}_{obs_{k}}$ and magnitude ($\beta^{(i)}_{obs_{k}}$) is proportional to how far the vehicle is away from the obstacle. $\mathbf{R}^{(i)}_{obs_{k}}$ is given as the cross product between $\mathbf{Z}$ and $\mathbf{X}^{(i)}_{obs_{k}}$ with a zero being appended to the third dimension of $\mathbf{X}^{(i)}_{obs_{k}}$ which originally lies in the 2D space. The vector resulting from the cross-product is perpendicular to $\mathbf{X}_{obs_{k}}^{(i)}$ which causes the vehicles to move in the clockwise direction around the obstacle as shown in Fig. \ref{fig:reference orientation near an obstacle}.

Likewise, the component for avoiding collision between the ego vehicle $i$ and another vehicle $j$ ($\mathbf{u}^{(i)}_{veh_{j}}$) is similar to Equation \ref{equ:reference orientation obstacle avoiding} except that the static obstacle radius $r_{obs}^{(k)}$ will be replaced by the other vehicle's radius $r_{veh}$ in the $\alpha^{(i)}_{veh_{j}}$ term. Secondly, the speed based dynamic margin is $|v^{(j)}_{t}| + |v^{(i)}_{t}| $ rather than just $|v^{(i)}_{t}|$. 

The overall collision avoiding component $\mathbf{u}^{(i)}_{coll}$ is given by:
\begin{equation}
 \mathbf{u}^{(i)}_{coll} = \sum_{k=1}^{N_{obs}} \mathbf{u}^{(i)}_{obs_{k}} + \sum_{j=1,j \ne i}^{N_{veh}} \mathbf{u}^{(i)}_{veh_{j}}
 \label{equ:reference orientation collision avoiding}
\end{equation}

Finally, the ideal reference orientation vector $\mathbf{\hat{u}}^{(i)}_{t+1}$ ($\mathbf{u_{ref}}$ in Fig. \ref{fig:pipeline overview}) is:
\begin{equation}
 \mathbf{\hat{u}}^{(i)}_{t+1} = \mathbf{f_{uni}}(\mathbf{u}^{(i)}_{tar} + \mathbf{u}^{(i)}_{coll}) 
 \label{equ:reference orientation}
\end{equation}

From this, the corresponding ideal reference orientation angle $\hat{\theta}^{(i)}_{t+1}$ for ego vehicle $i$ can be calculated by applying  $arctan2$ to $\mathbf{\hat{u}}^{(i)}_{t+1}$. However, kinematic constraints arising from the motion model (Equation \ref{equ:vehicle kinematic}) may prevent the vehicle from immediately attaining the ideal reference orientation
$\hat{\theta}^{(i)}_{t+1}$ in the next time step. Therefore, we instead use $\theta^{(i)}_{t+1}$ referred to as the \textit{real} orientation. It is the reachable orientation closest to  $\hat{\theta}^{(i)}_{t+1}$.  The unit vector corresponding to this real reference orientation angle $\theta^{(i)}_{t+1}$ for ego vehicle $i$ is $\mathbf{u}^{(i)}_{t+1}$ ($\mathbf{u_{real}}$ in Fig. \ref{fig:pipeline overview}) = $[\cos(\theta^{(i)}_{t+1}), \sin(\theta^{(i)}_{t+1})]^{T}$.

\subsection{Estimation of Reference Speed}\label{subsec:speed}

\begin{figure}[ht]
\centering
\subfigure[\centering \textbf{\label{fig:subfigure_a}}
{Forbidden Forward}]{{\includegraphics[width=0.49\linewidth]{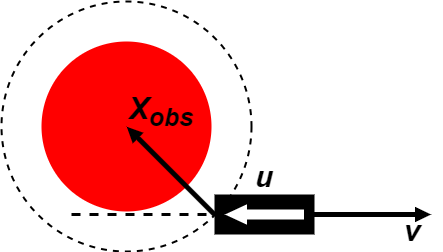}}}
\subfigure[\centering \label{fig:subfigure_b}
{Forbidden Backward}]{{\includegraphics[width=0.49\linewidth]{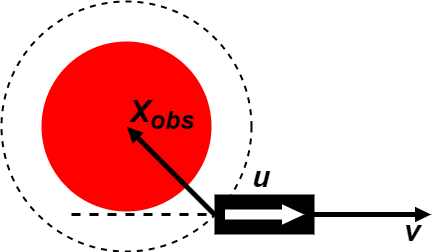}}}
\caption{Shows two different scenarios when the black ego vehicle is in a collision avoiding region. In Fig. \ref{fig:subfigure_a}, the ego vehicle is facing the obstacle, i.e. $\mathbf{u}^{T} \cdot \mathbf{X}_{obs}>0$, in which case the vehicle should be forbidden to move forwards. In Fig. \ref{fig:subfigure_b}, the ego vehicle is oriented away from the obstacle, i.e. $\mathbf{u}^{T} \cdot \mathbf{X}_{obs}<0$, in which case the vehicle should be prevented from moving backward.}
\label{fig:reference speed collision avoiding}
\end{figure}

The reference speed $v^{(i)}_{t+1}$ is chosen after determination of the reference orientation, which depends on the situation the vehicle is in. Fig. \ref{fig:reference speed collision avoiding} shows a scenario wherein a vehicle is at the same location next to an obstacle but with opposite orientations. This determines if the velocity should move the car forward or backward. 

We use the logical \emph{or} ($\vee$) and logical \emph{and}  ($\wedge$) operators \cite{mano2017digital} to describe the criteria for collision avoidance between ego vehicle $i$ and static obstacle $k$ under such situations as:
\begin{equation}
\begin{split}
F^{(i)}_{obs_{k}} & = (\alpha_{obs_{k}}^{(i)} + \epsilon_{c} \le 0) \wedge (\gamma_{obs_{k}}^{(i)} > 0) \\
B^{(i)}_{obs_{k}} & = (\alpha_{obs_{k}}^{(i)}  +\epsilon_{c} \le 0) \wedge (\gamma_{obs_{k}}^{(i)} < 0) \\
\gamma^{(i)}_{obs_{k}} & = {\mathbf{u}^{(i)}_{t+1}}^{T} \cdot \mathbf{X}^{(i)}_{obs_{k}} \\
\end{split}
\label{equ:reference speed obstacle avoiding}
\end{equation}

where $\alpha_{obs_{k}}^{(i)}$ is same as defined in Equation \ref{equ:reference orientation obstacle avoiding},  $\epsilon_{c}$ is the an additional tolerance for the collision checking region. $F^{(i)}_{obs_{k}}$ equalling $true$  indicates the ego vehicle $i$ is forbidden to drive forward towards obstacle $k$. This happens when the angle between the reference orientation ($\mathbf{{u}^{(i)}_{t+1}}$) and the vector from the ego vehicle to the obstacle ($\mathbf{X}^{(i)}_{obs_{k}}$) is less than $90^{\circ}$ i.e. $ \gamma^{(i)}_{obs_{k}}={\mathbf{u}^{(i)}_{t+1}}^{T} \cdot \mathbf{X}^{(i)}_{obs_{k}} > 0$. Likewise, $B^{(i)}_{obs_{k}}$ equaling $true$ happens when $ \gamma^{(i)}_{obs_{k}}={\mathbf{u}^{(i)}_{t+1}}^{T} \cdot \mathbf{X}^{(i)}_{obs_{k}} < 0$ indicating that the ego vehicle $i$ is forbidden to drive backward. 

Similarly for the case of ego vehicle $i$ and another vehicle $j$, the conditions for preventing the ego-vehicle from moving forward $F^{(i)}_{veh_{j}} $ or backward $B^{(i)}_{veh_{j}}$ are the same as described in Equation \ref{equ:reference speed obstacle avoiding} except that $\mathbf{X}^{(i)}_{veh_{j}}$ replaces $\mathbf{X}^{(i)}_{obs_{k}}$. 

Summarising the results above, the magnitude and sign of the velocity depends on the combination of these boolean variables i.e. $F^{(i)}_{obs_{k}}$, $F^{(i)}_{veh_{j}}, B^{(i)}_{obs_{k}}$ and $B^{(i)}_{veh_{j}}$ as follows:
\begin{equation}
\begin{split}
F^{(i)}_{coll} & = (\bigvee_{k=1}^{N_{obs}} F^{(i)}_{obs_{k}}) \vee (\bigvee_{j=1, j \ne i }^{N_{veh}} F^{(i)}_{veh_{j}} )\\
B^{(i)}_{coll} & = (\bigvee_{k=1}^{N_{obs}} B^{(i)}_{obs_{k}}) \vee (\bigvee_{j=1, j \ne i }^{N_{veh}} B^{(i)}_{veh_{j}}) \\
v^{(i)}_{coll}  & = 
 \begin{cases}
   v_{d} & B^{(i)}_{coll} \wedge (\neg F^{(i)}_{coll}) \\
   -v_{d} & (\neg B^{(i)}_{coll}) \wedge F^{(i)}_{coll} \\
   0 & B^{(i)}_{coll} \wedge F^{(i)}_{coll} \\
   v_{tar}^{(i)} & \text{otherwise} \\
 \end{cases}
\end{split}
\label{equ:reference speed collision avoiding}
\end{equation}
The first 3 conditions check whether or not there is a potential for collision. The velocity is $+v_{d}$, when the ego-vehicle is prevented from moving backwards ($B^{(i)}_{coll}$) but allowed to move forward ($\neg F^{(i)}_{coll}$) and $-v_{d}$ when vice versa. The reference velocity is zero when the ego-vehicle is prevented from moving both forward ($F^{(i)}_{coll}$) and backward ($B^{(i)}_{coll}$). In all other cases, the velocity is $v_{tar}^{(i)}$ defined as:
\begin{equation}
\begin{split}
v^{(i)}_{tar} & = 
\begin{cases}
   \xi^{(i)}_{p} \cdot \lambda^{(i)}_{p} \cdot v_{d} & \|\mathbf{X}^{(i)}_{tar}\|_{2} \le r_{p} \\
   v_{d} \cdot f_{sgn}(  {\mathbf{u}^{(i)}_{t+1}}^{T} \cdot \mathbf{\hat{u}}^{(i)}_{t+1}) & \text{otherwise} \\
\end{cases} \\
\lambda^{(i)}_{p} & =
\begin{cases}
\bar{\lambda}^{(i)}_{p} & (\|\mathbf{X}^{(i)}_{tar}\|_{2} < \epsilon_{p}) \wedge (|\theta^{(i)}_{tar}-\theta^{(i)}_{t+1}| < \epsilon_{o})  \\
  \sqrt{\bar{\lambda}^{(i)}_{p}} & \text{otherwise} \\
\end{cases} \\
\bar{\lambda}^{(i)}_{p} & = {minimum}(\frac{\|\mathbf{X}^{(i)}_{tar}\|_{2}}{r_{p}}+\frac{|\theta^{(i)}_{tar}-\theta^{(i)}_{t+1}|}{v_{d}}, 1) \\
\xi^{(i)}_{p} & = 
\begin{cases}
  1 & {\mathbf{u}^{(i)}_{t+1}}^{T} \cdot \mathbf{X}^{(i)}_{tar}>0.25 \\
  -1 & {\mathbf{u}^{(i)}_{t+1}}^{T} \cdot \mathbf{X}^{(i)}_{tar}<-0.25 \\
  f_{sgn}(\mathbf{u}^{(i)}_{t}) & \text{otherwise} \\
\end{cases} \\
\end{split}
\label{equ:reference speed parking}
\end{equation}

where $\epsilon_{p}$ and $\epsilon_{o}$ are the acceptable position and orientation tolerances for deciding to stop the vehicle at the target. When the ego vehicle is within the parking radius, it reduces its speed as it gets closer to the target position and orientation. This is achieved by the multiplier $\lambda^{(i)}_{p}$, which is proportional to $\bar{\lambda}^{(i)}_{p}$ when the vehicle state is very close to the target state and $\sqrt{\bar{\lambda}^{(i)}_{p}}$ when farther away from the tolerance. The square root accelerates the vehicle approaching its target when there is still some distance between the current state and target state, i.e. $\neg ((\|\mathbf{X}^{(i)}_{tar}\|_{2} < \epsilon_{p}) \wedge (|\theta^{(i)}_{tar}-\theta^{(i)}_{t+1}| < \epsilon_{o}))$. When the vehicle is very close to the target state, i.e. ($\|\mathbf{X}^{(i)}_{tar}\|_{2} < \epsilon_{p}) \wedge (|\theta^{(i)}_{tar}-\theta^{(i)}_{t+1}| < \epsilon_{o}$), the ratio $\bar{\lambda}^{(i)}_{p}$ prevents the vehicle from shaking forward and backward. For $\bar{\lambda}^{(i)}_{p}$, the first term: $\frac{1}{r_{p}} \cdot \|\mathbf{X}^{(i)}_{tar}\|_{2}$, decreases linearly with the ego vehicle distance to its target, and the second term, i.e. $\frac{1}{v_{d}} \cdot |\theta^{(i)}_{tar}-\theta^{(i)}_{t+1}|$, reduces linearly with the angle difference between the reference $\theta^{(i)}_{t+1}$ and target $\theta^{(i)}_{tar}$ orientation angles. However, we do not allow the reference parking speed $v_{tar}^{(i)}$ to be any higher than the default reference speed $v_{d}$. So, $\bar{\lambda}^{(i)}_{p}$ is clipped to a maximum of $1$. $\xi^{(i)}_{p}$ controls whether the ego vehicle moves forward or backward by checking ${\mathbf{u}^{(i)}_{t+1}}^{T} \cdot \mathbf{X}^{(i)}_{tar}$. Originally, the vehicle should move forward when facing the target, i.e. ${\mathbf{u}^{(i)}_{t+1}}^{T} \cdot \mathbf{X}^{(i)}_{tar}>0$, and move backward when backing towards the target, i.e.  ${\mathbf{u}^{(i)}_{t+1}}^{T} \cdot \mathbf{X}^{(i)}_{tar}<0$. However, to prevent the vehicle from changing direction at high frequency within short traveling distances, we set a margin allowing the vehicle to keep its previous direction when $|{\mathbf{u}^{(i)}_{t+1}}^{T} \cdot \mathbf{X}^{(i)}_{tar}| \le 0.25$. 

When the ego vehicle $i$ is out of the parking area of radius $r_{p}$, the reference speed $v_{t+1}^{(i)}$ takes the forn $v_{d} \cdot f_{sgn}(  {\mathbf{u}^{(i)}_{t+1}}^{T} \cdot \mathbf{\hat{u}}^{(i)}_{t+1})$, where the reference speed $v_{t+1}^{(i)}$ takes the default value $v_{d}$, but changes to negative, i.e. $-v_{d}$, when ${\mathbf{u}^{(i)}_{t+1}}^{T} \cdot {\mathbf{\hat{u}}^{(i)}_{t+1}}<0$.

The final ideal reference speed $\hat{v}^{(i)}_{t+1}$ for ego vehicle $i$ is:
\begin{equation}
    \hat{v}^{(i)}_{t+1} = v^{(i)}_{coll}
\label{equ:reference speed}
\end{equation}

Similar to the reference orientation, the ideal reference speed $\hat{v}^{(i)}_{t+1}$ may not achievable due to the limitation of the maximum pedal command. The real reference speed $v^{(i)}_{t+1}$ is therefore the reachable speed value closest to $\hat{v}^{(i)}_{t+1}$. \\

\noindent{\textbf{Calculation of Reference Steering Angle and Reference Pedal:}} Given the reference orientation and velocity, the Vehicle Kinematic Equation \ref{equ:vehicle kinematic} can be inverted to determine the reference steering angle $\varphi^{(i)}_{t}$ and pedal acceleration $p^{(i)}_{t}$ for controlling the vehicle at time $t$ :
\vspace*{-2mm}
\begin{equation}
\begin{split}
\varphi^{(i)}_{t} & = arctan2(\frac{(\theta^{(i)}_{t+1}-\theta^{(i)}_{t})}{v^{(i)}_{t} \cdot \gamma \cdot \Delta t}) \\
 p^{(i)}_{t} & = \frac{v^{(i)}_{t+1}- \beta \cdot v^{(i)}_{t}}{\Delta t}
\end{split}
\label{equ:reference steering angle and pedal}
\end{equation}

These reference control commands can either be directly used to control the vehicles, or as labels for training the GNN model in a self-supervised manner using the architecture of \cite{multiagent2023}. \\
\vspace*{-5mm}

\section{EXPERIMENTS} \label{sec:experiments}

\subsection{Algorithm Evaluation}\label{subsec:model_evaluation}

To assess the performance of our MA-DV$^{2}$F approach and its learning based counterpart (self-supervised GNN model), we compare with other recent learning and  search/optimization based algorithms in challenging, collision prone test cases. The recent learning based approaches include \cite{multiagent2023} which is a GNN model trained in a supervised manner and an extension of \cite{zhang2023gcbf} i.e. GCBF+ \cite{zhang2024gcbf+}, also a GNN model incorporating safety constraints using Control Barrier Functions (CBF).  Meanwhile, the two search/optimization based algorithms used for comparison include CL-MAPF \cite{wen2022cl} and CSDO \cite{10628993}. The test dataset comprises of challenging, collision prone scenarios with the number of vehicles ranging from 10 to 50 and static obstacles chosen between 0 and 25. Note that, all the static obstacles in the test cases are fixed to be circles with fixed radius because of the limitation of the CL-MAPF and CSDO environment setting and because it circumscribes an obstacle of arbitrary shape, allowing for safer behaviour. For the GCBF+, we use the DubinsCar model because it has vehicle kinematics most similar to ours. As GCBF+ has a completely different working domain with different map and agent sizes, we scale our test cases when running GCBF+ under the rule that the ratio of the map to agent size remains equal across all the algorithms. Details regarding the generation of test samples are given in the supplementary file on the project page. 

Table \ref{table:success rate of different models} shows the evaluation results of the different methods across the different vehicle-obstacle combinations described earlier against the \textit{success rate} metric \cite{zhang2024gcbf+}. It measures the percentage of vehicles that successfully reach their targets within a tolerance without colliding with other objects along the way. The results show that our MA-DV$^{2}$F outperforms other algorithms achieving almost 100\% success rate across all vehicle-obstacle combinations. Even the self-supervised GNN model performs far better than other algorithms when scaling the number of vehicles. Only the CSDO algorithm has slightly better results than our GNN model when the number of agents are low (20). However, CSDO's performance drops drastically as the number of agents are increased under these challenging conditions. Note that the GCBF+ pipeline only checks whether the agents reach their target positions but ignores the target orientations as shown in the project page: \href{https://yininghase.github.io/MA-DV2F/\#MC}{https://yininghase.github.io/MA-DV2F/\#MC} which explains why it has such poor performance. Therefore, we additionally evaluate GCBF+ by ignoring the orientation and only considering the position. Results of which are shown in the second last column. Even then, the GCBF+, does not match the performance of MA-DV$^{2}$F.

\begin{table*}[!ht]
\centering
\resizebox{\textwidth}{!}{
\begin{tabular}{||c c c c c c c c c||} 
\hline
\hline
Number of & Number of & MA-DV$^{2}$F & Self-Supervised  & CSDO & CL-MAPF & GCBF+ & GCBF+ & Supervised  \\
Vehicles & Obstacles & (Ours) & Model (Ours) & \cite{10628993} &  \cite{wen2022cl} & \cite{zhang2024gcbf+} & (only position) & Model \cite{multiagent2023} \\ [0.5ex]
\hline
\multicolumn{9}{||c||}{success rate $\uparrow$ } \\ 
\hline
10 & 0 & \textbf{1.0000} &  0.9929 & 0.9693 & 0.4290 & 0.0613 & 0.9021 & 0.7285 \\
\hline
20 & 0 & \textbf{1.0000} & 0.9895 & 0.9400 & 0.4963 & 0.0563 & 0.8169 & 0.2041 \\
\hline
30 & 0 & \textbf{1.0000} & 0.9793 & 0.8820 & 0.4977 & 0.0458 & 0.7550 & 0.0477 \\
\hline
40 & 0 & \textbf{1.0000} & 0.9760 & 0.7063 & 0.5632 & 0.0451 & 0.7080 & 0.0137 \\
\hline
50 & 0 & \textbf{1.0000} & 0.9686 & 0.6523 & 0.5992 & 0.0426 & 0.6618 & 0.0045 \\
\hline
10 & 25 & \textbf{0.9952} & 0.9458 & 0.9682 & 0.5640 & 0.0509 & 0.7192 & 0.3734 \\
\hline
20 & 25 & \textbf{0.9902} & 0.9208 & 0.9663 & 0.5997 & 0.0393 & 0.6479 & 0.0756 \\
\hline
30 & 25 & \textbf{0.9844} & \textit{0.8998} & 0.8456 & 0.6320 & 0.0385 & 0.5754 & 0.0196 \\
\hline
40 & 25 & \textbf{0.9772} & \textit{0.8723} & 0.6723 & 0.6531 & 0.0330 & 0.5211 & 0.0068 \\
\hline
50 & 25 &\textbf{0.9704} & 0.8504 & 0.5897 & 0.6749 & 0.0290 & 0.4789 & 0.0028 \\
\hline
\end{tabular}
}
\caption{Shows the \textit{Success} rate metric (Higher is better) for different models, i.e. our MA-DV$^{2}$F, our self-supervised GNN model, CSDO \cite{10628993}, CL-MAPF \cite{wen2022cl}, GCBF+ \cite{zhang2024gcbf+} and the supervised GNN model \cite{multiagent2023}. As GCBF+ does not consider vehicle orientation when reaching the target, we additionally evaluate the \textit{Success} rate by only checking vehicle position and disregarding the orientation. For each row, the bold number show the highest success rate among all algorithms.}
\label{table:success rate of different models}
\vspace{-1.0em}
\end{table*}

\subsection{Discussion}

\begin{figure*}[ht]
\centering
    \subfigure[\centering{safe rate with no obstacles}]{{\includegraphics[width=0.22\textwidth]{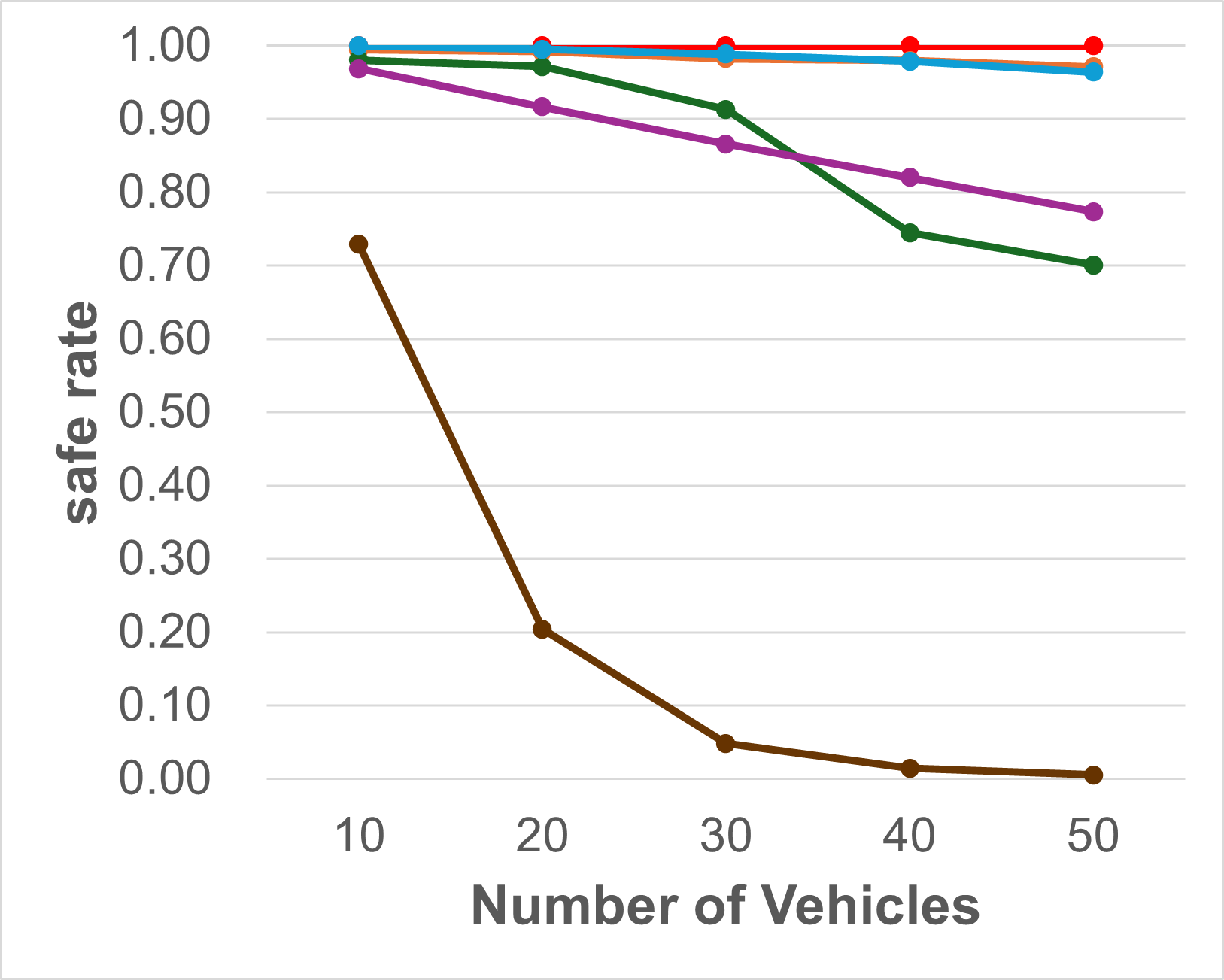}}}
    \subfigure[\centering{safe rate with 25 obstacles}]{{\includegraphics[width=0.22\textwidth]{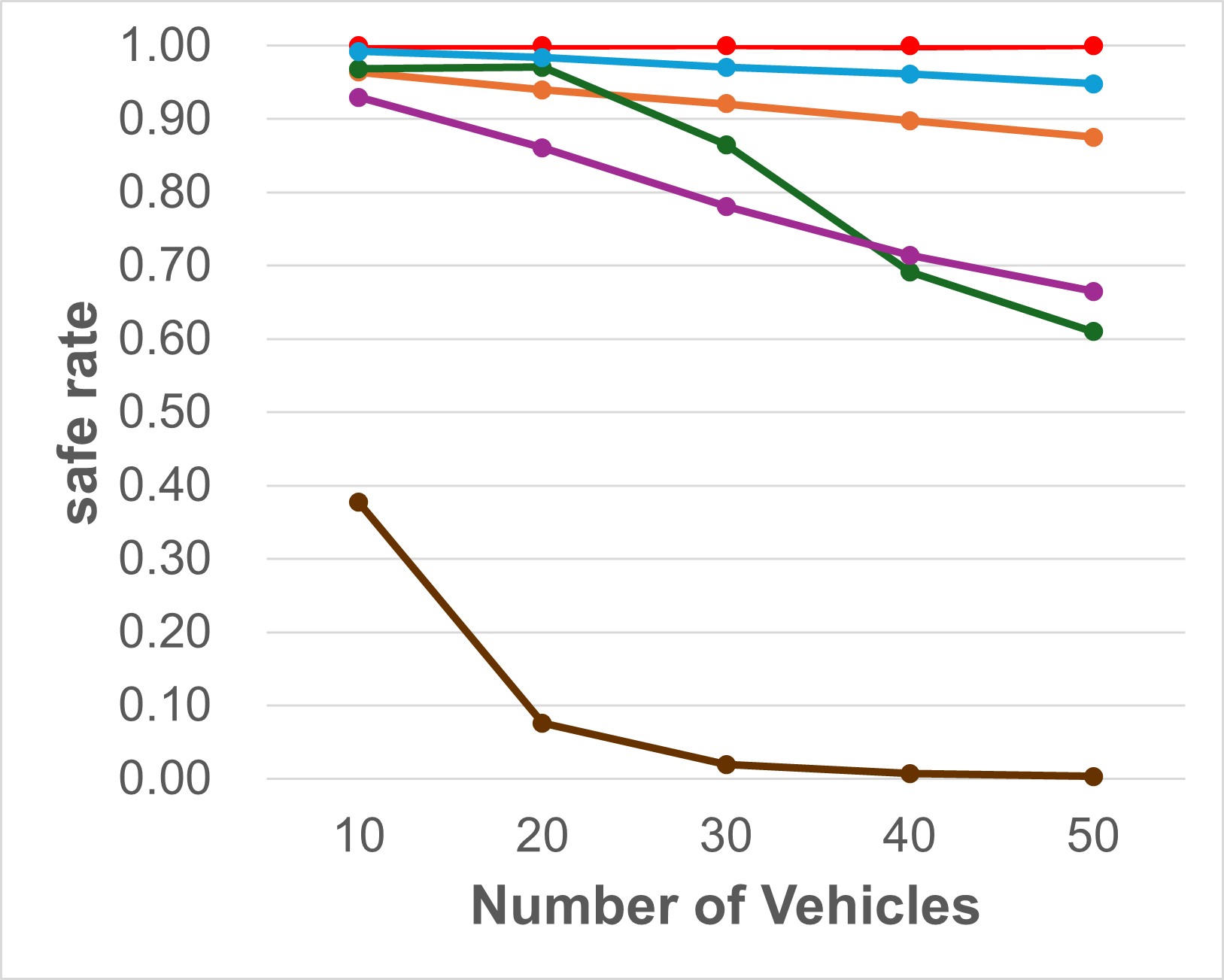}}}
    \subfigure[\centering{reach rate with no obstacles}]{{\includegraphics[width=0.22\textwidth]{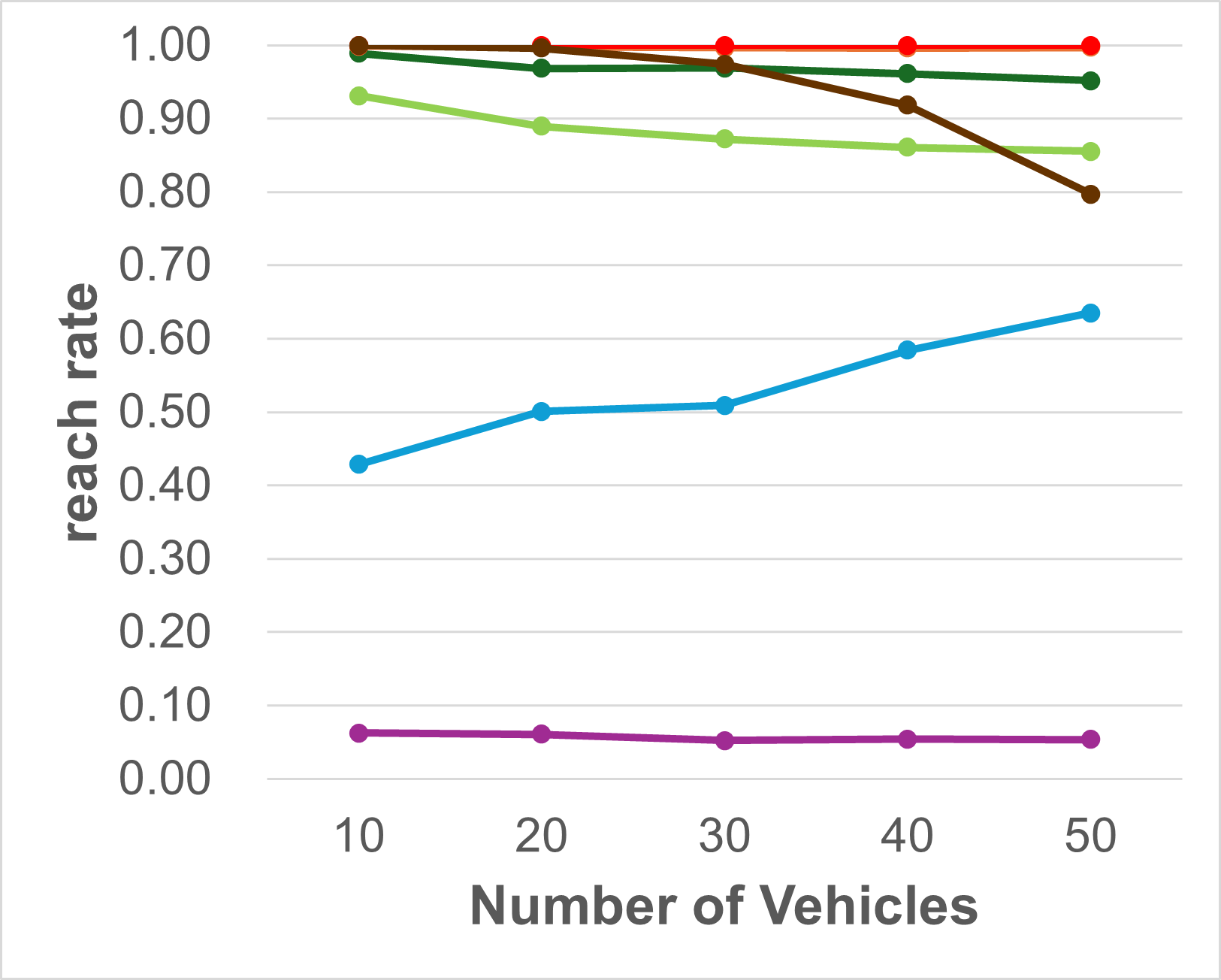}}}
    \subfigure[\centering{reach rate with 25 obstacles}]{{\includegraphics[width=0.22\textwidth]{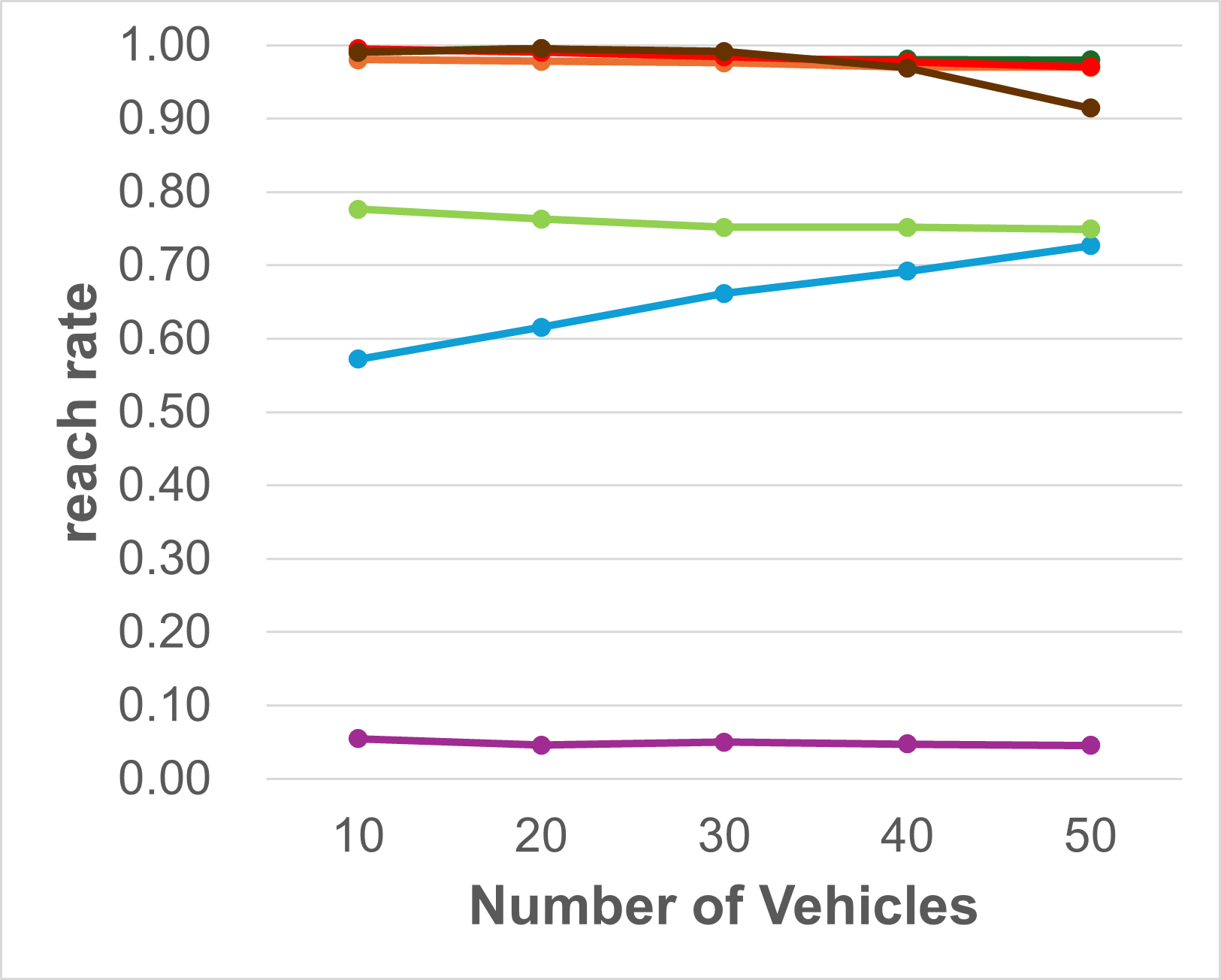}}}
    \subfigure[\centering{Legend}]{{\includegraphics[width=0.08\textwidth]{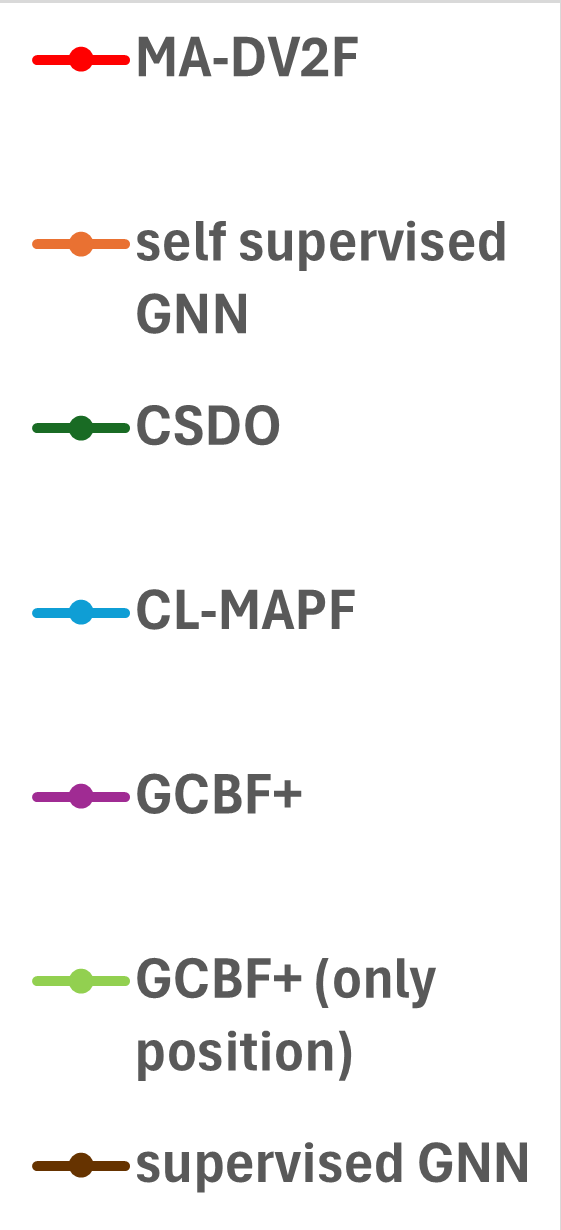}}}
    \caption{Shows the \textit{Safe} and \textit{Reach} rate metrics (Higer is better) for the different models, i.e. our MA-DV$^{2}$F, our self-supervised GNN model, supervised GNN model \cite{multiagent2023}, CSDO \cite{10628993}, CL-MAPF \cite{wen2022cl} and GCBF+ \cite{zhang2024gcbf+}.}
    \label{fig:safe success rate of different models}
\end{figure*}

\noindent{\textbf{Investigating failure Causes:}} Note that the \textit{Success rate} metric measures a model's ability to not only reach its destination but also avoid collisions.  
Therefore, for challenging test cases, a navigation algorithm may fail due to two main reasons: the algorithms behave too aggressively by driving the vehicles to their targets albeit at the cost of colliding with other agents or behaves too conservatively to ensure safety of the vehicles resulting in some vehicles getting stuck mid-way without reaching their targets. Therefore to disambiguate between the two causes, we additionally evaluate all algorithms against the \textit{Reach rate} and \textit{Safe rate}  metrics as proposed in \cite{zhang2024gcbf+}. \textit{Reach rate} measures the percentage of vehicles that reach their targets within a tolerance disregarding any collisions along the way. Meanwhile, the \textit{Safe rate} calculates the percentage of vehicles that do not collide with any other agents irrespective of whether or not they reach their targets. Fig. \ref{fig:safe success rate of different models} shows the performance of the different methods against the  \textit{Reach Rate} and \textit{Safe Rate}  metrics. It can be seen that the supervised GNN model behaves rather aggressively reaching the target in majority of the cases albeit at the cost of multiple collisions leading to high \textit{Reach} but low \textit{Safe} rates. In contrast, CL-MAPF \cite{wen2022cl} takes a greedy strategy in its optimization/search pipeline. It can quickly find the path for some vehicles which are easy to plan. But then it is unable to find paths for other vehicles and gets stuck in this partial solution. This greedy strategy lead to a low reach rate, but high safe rate since no collisions happen among vehicles that do not move.  \\
GCBF+ has a higher safe rate than reach rate. This is not because the vehicles fail to reach their target, but rather because they reach the target at an incorrect orientation. This is aligned with the intuition described in \ref{subsec:model_evaluation} and is further corroborated by the fact that when orientation is ignored in the evaluation and only position is considered, the reach rate jumps drastically. Nevertheless, it is still much lower than both our MA-DV$^{2}$F and its GNN self-supervised counterpart. They are the only 2 methods which maintain a consistent high performance against both metrics across all vehicle-obstacle combinations. \\

\noindent{\textbf{Preventing Bottlenecks:}} A common problem with other planning algorithms is that vehicles tend to crowd within a small sub-region in the map. This leads to these algorithms either unable to find an optimal path resulting in all vehicles becoming stationary at their place (low reach rate) or finding sub-optimal paths resulting in more collisions among vehicles (low safe rate). To prevent such bottlenecks in our MA-DV$^2$F model, Equation \ref{equ:reference orientation obstacle avoiding} causes all the vehicles to drive in the clockwise direction when encountering other agents. This leads to a roundabout driving behavior which breaks the bottleneck and can be visualized on the project page: \href{https://yininghase.github.io/MA-DV2F/\#RE}{https://yininghase.github.io/MA-DV2F/\#RE}. Due to this, the vehicles are capable of eventually reaching their targets,by making a detour around the other agents, resulting in both a high \textit{Reach} and \textit{Safe} rate. Our trained GNN model also adapts this behaviour.\\

\noindent{\textbf{Intermediate Success Rate:}} One might wonder if the MA-DV$^2$F method outperforms every other method, what is the advantage of its self-supervised GNN model counterpart? One reason is that MA-DV$^2$F behaves conservatively in some simple test cases even though there is no collision risk nearby, causing the vehicles to take a longer time to finish the journey. This is because the speed is limited to $v_{d}$. On the other hand, the self-supervised GNN counterpart is free from this restriction. If can move faster when it is far away from its target and there is less risk of collision with other objects. The figures in project page: \href{https://yininghase.github.io/MA-DV2F/\#ISR}{https://yininghase.github.io/MA-DV2F/\#ISR}  show the difference of the success rate between the self-supervised GNN model and MA-DV$^2$F as a function of time. At the beginning when the vehicles tend to be far away from their targets, the GNN allows high velocity for vehicles, thereby causing some vehicles to reach their target faster leading to a success rate greater than that for MA-DV$^2$F. This can also seen in the example in project page: \href{https://yininghase.github.io/MA-DV2F/\#MC}{https://yininghase.github.io/MA-DV2F/\#MC}, the vehicles driven by a GNN can drive faster towards the targets than that by MA-DV$^2$F. However, maintaining a high velocity also leads to higher risk of collision when encountering challenging situations as time progresses and the success rate of MA-DV$^2$F will gradually catch up and  eventually exceed the GNN.

\begin{table}[!ht]
\centering
\resizebox{\linewidth}{!}{
\begin{tabular}{||c c c c c||} 
\hline
\hline
Number of & Number of & MA-DV$^2$F & CSDO & CL-MAPF \\
Vehicles & Obstacles & (Ours) & \cite{10628993} &  \cite{wen2022cl} \\ [0.5ex]
\hline
\multicolumn{5}{||c||}{total runtime for 1000 test cases $\downarrow$ } \\ 
\hline
10 & 0 & \textit{\textbf{4.152}} & 195.919 & 125643.851 \\
\hline
20 & 0 & \textit{\textbf{5.848}} & 1044.360 & 219251.537 \\
\hline
30 & 0 & \textit{\textbf{7.885}} & 6179.536 & 328471.949 \\
\hline
40 & 0 & \textit{\textbf{8.492}} & 12342.014 & 376417.421 \\
\hline
50 & 0 & \textit{\textbf{10.512}} & 18306.779 & 395962.103 \\
\hline
10 & 25 & \textit{\textbf{10.322}} & 352.096 & 89838.921 \\
\hline
20 & 25 & \textit{\textbf{12.260}} & 2213.838 & 163001.473 \\
\hline
30 & 25 & \textit{\textbf{13.733}} & 9148.955 & 190907.288 \\
\hline
40 & 25 & \textit{\textbf{15.562}} & 14350.735 & 224962.016 \\
\hline
50 & 25 & \textit{\textbf{18.405}} & 16598.068 & 253495.187 \\
\hline
\end{tabular}
}
\caption{Shows the total runtime in seconds (Lower is better) for the 1000 test cases for each vehicle-obstalce combination of the different models, i.e. our MA-DV$^2$F, CSDO \cite{10628993} and CL-MAPF \cite{wen2022cl}.}
\label{table:runtime of different models}
\end{table}

\noindent{\textbf{Runtime Analysis:}} We compare the runtime of MA-DV$^{2}$F with concurrent search based methods (CSDO and CL-MAPF). TABLE \ref{table:runtime of different models} shows the total runtime of all the 1000 test cases for each vehicle-obstalce combinations. All methods were evaluated on a machine with an Intel Core i7-10750H CPU and GeForce RTX 2070 GPU. As can be seen, our MA-DV$^2$F, is orders of magnitude faster than its peers, since it has the ability to run the different scenarios within a batch in parallel. Meanwhile CSDO and CL-MAPF are search/optimization-based algorithms that need to solve each test cases one-by-one. Note that CL-MAPF would continue its search until a solution is found, which is why the maximum allowed runtime needs to be clipped, otherwise the runtime would be even larger. Note that the evaluations in the experiments were done for only upto 50 vehicles, since other algorithms are either extremely slow or have drastically reduced performance when scaling. However, our method is capable of scaling to a far greater number of vehicles than just 50 as can be observed on the project page for scenarios with 100, 125, 250 vehicle-obstacle combinations: \href{https://yininghase.github.io/MA-DV2F/\#LE}{https://yininghase.github.io/MA-DV2F/\#LE}.

Lastly, note that the runtime analysis is not done for the learning based methods since, it is dependent on the GPU rather than the algorithm itself. Therefore, for the same model architecture, the runtime will be the same for all learning based algorithms.\\

\noindent{\textbf{Limitations:}} If the vehicles are densely packed or their targets are within the safety margins of one another, then due to their non-holonomic behavior there might not be enough space for them to navigate without risking a collision. In such a scenario, the vehicles will act conservatively and hesitate to proceed so as to avoid collisions leading to less vehicles reaching the target. A similar outcome is observed, if some vehicles start behaving unexpectedly, wherein the safety margin would need to be increased to mitigate the risk of collision, albeit at the expense of reaching the target. More details on  sensitivity analysis of the effect of change in static component of the safety margin ($r_{c}$) and visualization of the limitations is in the supplementary file and the project page.
\section{Conclusion}\label{sec:conclusion}

This work introduced MA-DV$^{2}$F, an efficient algorithm for multi-vehicle navigation using dynamic velocity vector fields. Experimental results showed that our approach seamlessly scales with the number of vehicles. When compared with other concurrent learning and search based methods,  MA-DV$^{2}$F has higher success rate, lower collision metrics and higher computational efficiency.

\noindent{\textbf{\textit{Acknowledgements:}}} We thank Ang Li for the discussions and his feedback on the setup of the experiments.
\bibliographystyle{IEEEtran}
\bibliography{root}
\clearpage
\section{Supplementary}\label{sec:supplementary}
\subsection{Further Explanation of MA-DV$^{2}$F}

In this Subsection, we further elaborate some the design choices for Equation \ref{equ:reference orientation target approaching} used to determine the reference orientation. The Equation is repeated here again for clarity:
\begin{equation}
\begin{split}
    \mathbf{u}^{(i)}_{tar} & =
    \begin{cases}
      \mathbf{f_{uni}}(\mathbf{X}^{(i)}_{tar}) \cdot \xi_{tar}^{(i)} & \|\mathbf{X}^{(i)}_{tar}\|_2 > r_{p} \\
     \mathbf{f_{uni}}(\mathbf{U}^{(i)}_{tar} +  \lambda^{(i)}_{tar}  \cdot  \mathbf{f_{uni}}(\mathbf{X}^{(i)}_{tar})) & \text{otherwise}
    \end{cases} \\ 
  \lambda^{(i)}_{tar} & = (\frac{\|\mathbf{X}^{(i)}_{tar}\|_2}{r_{p}} + f_{pos}(\|\mathbf{X}^{(i)}_{tar}\|_2-\epsilon_{p})) \cdot f_{sgn}({\mathbf{X}_{tar}^{(i)}}^{T} \cdot \mathbf{U}^{(i)}_{tar})\\
  \xi_{tar}^{(i)} & = 
  \begin{cases}
    1 & \|\mathbf{X}^{(i)}_{tar}\|_2 \ge 0.5 \cdot v^{2}_{d} + r_{p} \\
    f_{sgn}({\mathbf{X}_{tar}^{(i)}}^{T} \cdot \mathbf{U}^{(i)}_{t}) & \text{otherwise}\\
  \end{cases}\\
\end{split}
\nonumber
\end{equation}

\begin{figure}[ht]
\centering
    \subfigure[\centering \label{fig:subfigure_5a} {Keep Reference Orientation}]{{\includegraphics[width=0.49\linewidth]{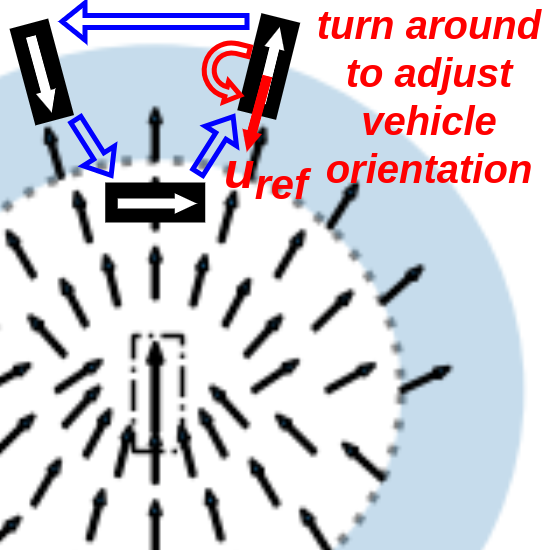}}}
    \subfigure[\centering \label{fig:subfigure_5b} {Reverse Reference Orientation}]{{\includegraphics[width=0.49\linewidth]{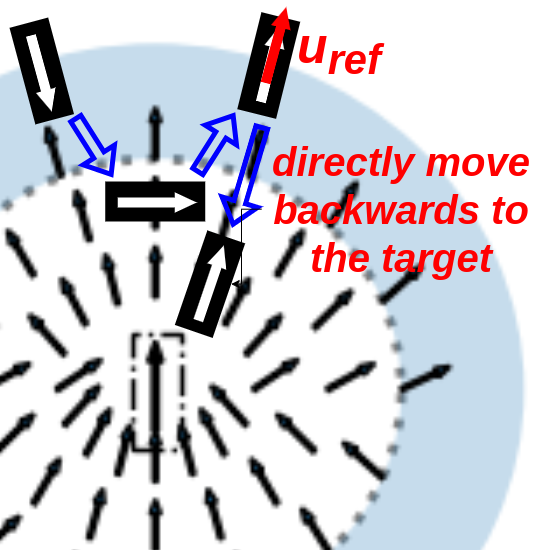}}}
    \caption{Shows the two different circumstances of the ego-vehicle overshooting its target and entering the marginal parking region (shaded blue region). In Fig. \ref{fig:subfigure_5a}, The reference orientation vector $\mathbf{u}_{ref} = \mathbf{f_{uni}}(\mathbf{X}^{(i)}_{tar})$ points towards the target, while the current vehicle orientation is opposite to the reference orientation. Thus, the vehicle needs to turn around again to match the reference orientation. In Fig. \ref{fig:subfigure_5b}, the reference orientation vector $\mathbf{u}_{ref} = - \mathbf{f_{uni}}(\mathbf{X}^{(i)}_{tar})$ is flipped based on the current vehicle orientation. In this case, the vehicle does not need to turn around but directly move backwards to the target.}
    \label{fig:reference orientation in gray region}
    \vspace{+2.0em}
\end{figure}

\noindent{\textbf{Circular Motion Prevention:}} Equation \ref{equ:reference orientation target approaching} shows that when the ego-vehicle $i$ is far away from the parking threshold, the reference orientation is in the direction of $\mathbf{X}^{(i)}_{tar}$. However, as the ego-vehicle approaches the target and the velocity is high, then the vehicle might overshoot the target and enter the shaded marginal parking threshold region causing the current ego-vehicle orientation $\mathbf{U}^{(i)}_{t}$ to be opposite to $\mathbf{X}^{(i)}_{tar}$ as seen in Fig. \ref{fig:subfigure_5a}. This will make the vehicle go in circles in an attempt to align itself with the reference orientation. A better alternative would be switch the reference orientation direction so that it is aligned with the ego-vehicle orientation and drive the car in reverse as shown in Fig. \ref{fig:subfigure_5b}. This is done by dynamically switching the reference orientation depending on the sign of the dot product: $f_{sgn}({\mathbf{X}_{tar}^{(i)}}^{T}. \mathbf{U}^{(i)}_{t})$. 

\begin{figure}[ht]
\centering
    \subfigure[\centering \label{fig:subfigure_6a}
    {Without Refining Item}]{{\includegraphics[width=0.49\linewidth]{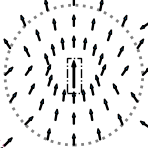}}}
    \subfigure[\centering \label{fig:subfigure_6b}
    {With Refining Item}]{{\includegraphics[width=0.49\linewidth]{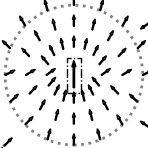}}}
    \caption{Shows the reference orientations near the target with/without the refining term $f_{pos}(\|\mathbf{X}^{(i)}_{tar}\|_2-\epsilon_{p})$ in $\lambda^{(i)}_{tar}$ in Equation \ref{equ:reference orientation target approaching}. Fig. \ref{fig:subfigure_6a} shows the situation without the refining term. The reference direction arrows in the left side or right side of the target position are more parallel to the target direction. This will lead to the vehicle shaking forwards and backwards there rather than approaching the target. Therefore, the refining item $f_{pos}(\|\mathbf{X}^{(i)}_{tar}\|_2-\epsilon_{p})$ is added to $\lambda^{(i)}_{tar}$. The new reference direction arrows shown in Fig. \ref{fig:subfigure_6b} have more biases towards the target position, which can accelerate the parking process. }
    \label{fig:reference orientation at target}
\end{figure}

\noindent{\textbf{Parking behaviour Refinement:}} In Equation \ref{equ:reference orientation target approaching}, an additional term: $f_{pos}(\|\mathbf{X}^{(i)}_{tar}\|_2-\epsilon_{p})$ was introduced in $\lambda^{(i)}_{tar}$ to refine the parking behavior when the vehicle position is exactly on either the left or the right side of the target.
As can be seen in Fig. \ref{fig:reference orientation at target}, this additional refinement term causes the reference orientation to be more biased towards the target position within the parking region. The ego vehicle aligns itself better to the reference orientation there, while simultaneously allowing it to  reach the final target quicker. Removing this term would make the reference orientations rather parallel to the final target causing the ego vehicle to oscillate forward and backwards around the target leading to a longer parking time. This is particularly true  when the ego vehicle position is exactly to the left or to the right side of the target. 

\subsection{Self-supervised Training of GNN Model} \label{sec:self supervised learning}

\begin{figure}[!ht]
\centering
\includegraphics[width=\linewidth]{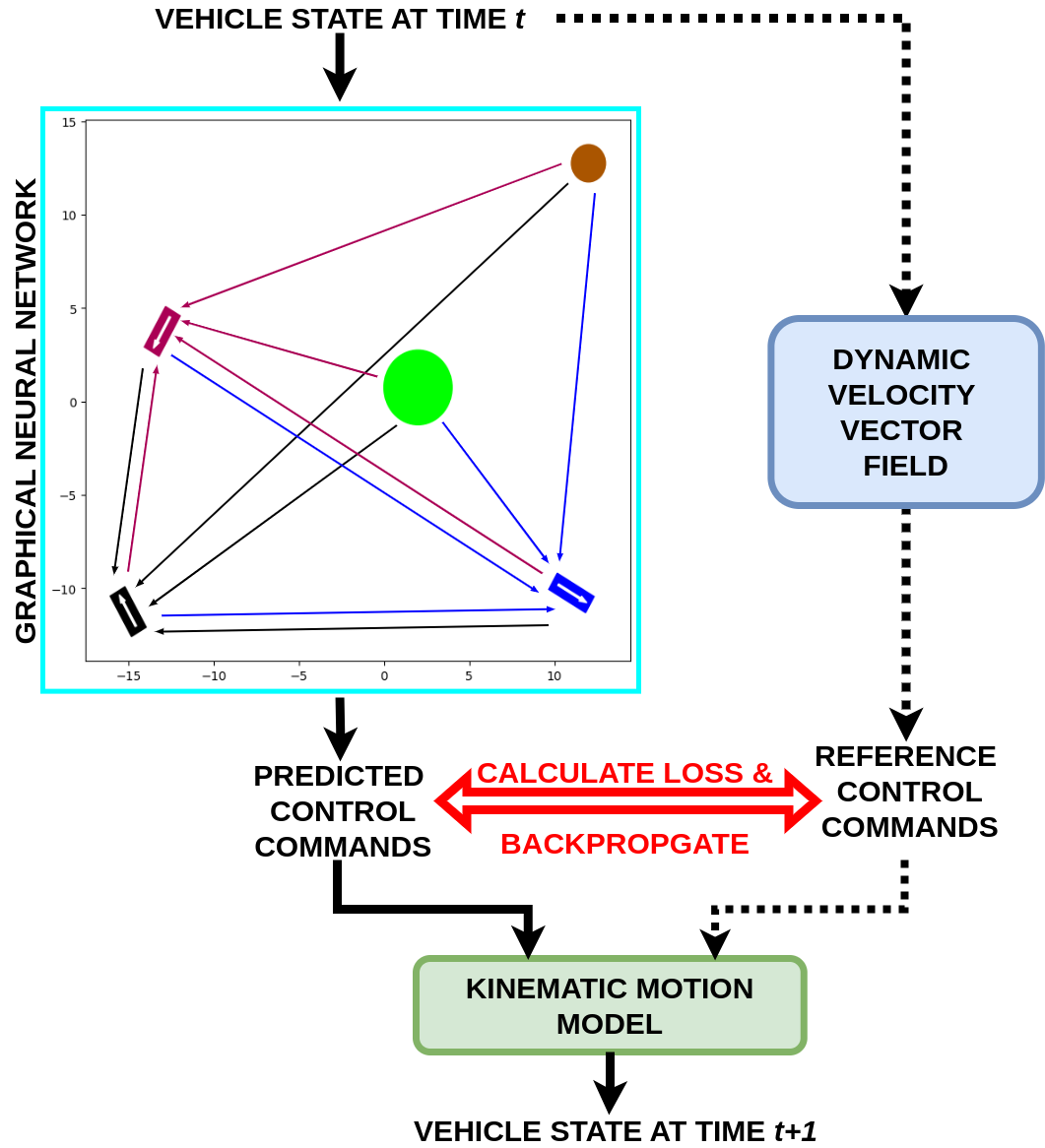}
\caption{Shows the pipeline of the self-supervised learning part. The model is trained by running simulation on different training cases without existing labels. For the simulation with $T$ steps in total, at each step of the simulation, the DV$^{2}$F are calculated for all the vehicles. For each vehicle, its DV$^{2}$F gives the reference control variables as training labels. At the same time, the state of the vehicles and the obstacles will be inputted to the training model, obtaining the predicted control variables from the model. Then, the loss is calculated based on the reference and predicted control variables and back-propagated to update the weights of the model. The predicted control variables will also be passed to the vehicle kinetic equation to update the vehicle states for the next simulation step. This process iterates until finishing this simulation turn.}
\label{fig:self supervised training pipeline}
\end{figure}

The reference steering angles $\varphi^{(i)}_{t}$ and reference pedal acceleration $p^{(i)}_{t}$  mentioned in Section \ref{sec:framework} cannot only be used to control the vehicles directly, but also as labels to supervise training of the GNN model. This approach can therefore additionally be used to train a learning based Graphical Neural Network (GNN) controller in a self-supervised manner using these control labels and network architecture given in \cite{multiagent2023}. However, \cite{multiagent2023} requires  running an optimization based procedure offline to collect enough training data with ground truth labels. This is a computationally expensive and slow process when the number of agents are large. In contrast, our self-supervised learning method is capable of directly training the model online during the simulation process without collecting ground truth labels beforehand. Note that in \cite{multiagent2023}, all the vehicle nodes in the GNN model will have incoming edges from any other vehicle or obstacle nodes. In this case, the number of edges in the graph will grow quadratic to the number of agents(neighbouring vehicles/obstacles), leading to a heavy computational burden when when scaling to more agents. Thus, we remove edges between the neighboring agent $j$ or $k$ and the ego vehicle $i$ if the distance $\|X^{(i)}_{j}\|$ between them is greater than the threshold given by: 
\begin{equation}
D^{(i)}_{j / k} = 
\begin{cases}
2 \cdot r_{veh}+|v^{(i)}_{t}|+|v^{(j)}_{t}|+2\cdot r_{c} & \text{$j$ is a vehicle} \\
r^{(k)}_{obs}+r_{veh}+|v^{(i)}_{t}|+2\cdot r_{c} & \text{$k$ is an obstacle} \\
\end{cases}
\label{equ:edge filtering threshold}
\end{equation} \\

\noindent{\textbf{Training Pipeline:}} The self-supervised learning pipeline is shown in Fig. \ref{fig:self supervised training pipeline}. At the start of training, different training samples with multiple agents placed at random positions are generated as the start points of the DV$^2$F simulations. Each sample only contains the states of the vehicles and static obstacles without any control labels. At each step during the simulation, the reference control variables of each vehicle is determined online using the DV$^2$F according to the states of the vehicles and obstacles at that time. The loss is calculated on the fly based on the reference control variables and the predicted control variables by the model.  This loss is then back-propagated to update the model immediately at the current step. Unlike \cite{multiagent2023} which solves the individual optimization problem for each case one-by-one, our dynamic velocity vector field determines reference control variables in a closed-form solution. Thus, we can run one simulation with multiple cases as a batch running in parallel, and then change to another batch for the next turn of simulation until finishing all the training cases as an epoch. 

\noindent{\textbf{Loss Function:}} The dynamic velocity vector field gives a low reference speed limited by $|v_{d}|$. However, if the vehicle is still far away from its target and has low risk to colliding with other agents, the speed limit of $|v_{d}|$ can be removed allowing the vehicle to move faster to its target. To this end,
we first define a vehicle state cost to evaluate vehicle control. Assume the current vehicle position $(x^{(i)}_{t},y^{(i)}_{t})$, orientation $\theta^{(i)}_{t}$ and speed $v^{(i)}_{t}$ are fixed, for the given  vehicle control variables $\varphi^{(i)}_{t}$ and $p^{(i)}_{t}$, we first apply the vehicle kinematic equation to obtain $x^{(i)}_{t+2}$, $y^{(i)}_{t+2}$, $\theta^{(i)}_{t+1}$ and $v^{(i)}_{t+1}$. We redefine $\|\mathbf{X}^{(i)}_{tar}\|$, $\|\mathbf{X}^{(i)}_{obs_{k}}\|$ and $\|\mathbf{X}^{(i)}_{veh_{j}}\|$ for the time $t+2$.  The vehicle state cost $C(\varphi^{(i)}_{t}, p^{(i)}_{t})$ is then calculated as follows:
\begin{equation}
\begin{split}
C & = C_{tar} + C_{obs} + C_{veh}\\
C_{tar} & = \|\mathbf{X}^{(i)}_{tar}\|_{2} \\
C_{obs} & = \sum_{k=1}^{N_{obs}} f^{2}_{max}(-\alpha^{(i)}_{obs_{k}},0)+f_{max}(-\alpha^{(i)}_{obs_{k}},0) \\
C_{veh} & = \sum_{j=1, j \ne i}^{N_{veh}} f^{2}_{max}(-\alpha^{(i)}_{veh_{j}},0)+f_{max}(-\alpha^{(i)}_{veh_{j}},0) \\
\end{split}
\label{equ: vehicle state cost}
\end{equation}

where $\alpha_{obs_{k}}^{(i)} = \|\mathbf{X}^{(i)}_{obs_{k}}\|_2-r_{obs}^{(k)}-r_{veh}-(r_{c} + |v^{(i)}_{t}|)$ and $\alpha_{veh_{j}}^{(i)} = \|\mathbf{X}^{(i)}_{veh_{j}}\|_2-2 \cdot r_{veh}-(r_{c} + |v^{(i)}_{t}|)$ are defined same as in Equation \ref{equ:reference orientation obstacle avoiding} mentioned in Section \ref{subsec:orientation}, $C_{tar}$ measures the distance of the ego vehicle $i$ to its target, and $C_{obs}$ and $C_{veh}$  evaluate the collision risk of the ego vehicle. The gradient of this vehicle state cost should alone be enough to guide the GNN model in learning to reach the target while avoiding collisions. However, in practice, the GNN does not converge to the optimal due to high non-linearities in Equation \ref{equ: vehicle state cost}. Therefore, we combine this equation with the labels obtained from the dynamic velocity vector field which expedites the model training. 

The overall loss function used in this pipeline is defined as follows:
\begin{equation}
\begin{split}
L & = L_{steer} + L_{pedal} \\
L_{steer} & = (\varphi^{(i)}_{t} - \Tilde{\varphi}^{(i)}_{t})^{2} \\ 
L_{pedal} & =
\begin{cases}
 \Delta C + f_{pos}(\Delta C)\cdot (\Delta C) ^{2} & \alpha^{(i)}_{tar} \wedge \beta^{(i)}_{tar}\\
  (p^{(i)}_{t} - \Tilde{p}^{(i)}_{t})^{2}  & \text{otherwise}\\
\end{cases} \\
\Delta C & = C(\varphi^{(i)}_{t}, \Tilde{p}^{(i)}_{t})
 - C(\varphi^{(i)}_{t}, p^{(i)}_{t})\\
\alpha^{(i)}_{tar} & = \|\mathbf{X}^{(i)}_{tar}\|_{2} - |\Tilde{v}^{(i)}_{t+1}| - r_{p}>0 \\
\beta^{(i)}_{tar} & =  (|\Tilde{v}^{(i)}_{t+1}| > v_{d}) \wedge (|v^{(i)}_{t+1}| = v_{d}) \wedge (\Tilde{v}^{(i)}_{t+1} \cdot v^{(i)}_{t+1}>0)\\
\end{split}
\label{equ: self supervised loss}
\end{equation}

where the $\varphi^{(i)}_{t}$, $p^{(i)}_{t}$ and $v^{(i)}_{t+1}$ are the reference steering angle, reference pedal and calculated reference speed from dynamic velocity vector field, the $\Tilde{\varphi}^{(i)}_{t}$, $\Tilde{p}^{(i)}_{t}$ and $\Tilde{v}^{(i)}_{t+1}$ are the corresponding values predicted by the GNN model. 
During training, the steering angle $\Tilde{\varphi}^{(i)}_{t}$ is fully supervised by reference steering angle $\varphi^{(i)}_{t}$. However, the pedal acceleration loss comprises one of two parts depending on the condition of the vehicle. If the vehicle is far way from the parking region, i.e. $\|\mathbf{X}^{(i)}_{tar}\|_{2} - |\Tilde{v}_{t+1}| - r_{p}>0$, and the default reference speed $v_{d}$ limits the predicted speed, i.e. $(|\Tilde{v}_{t+1}| > v_{d}) \wedge (|v_{t+1}| = v_{d}) \wedge (\Tilde{v}_{t+1} \cdot v_{t+1}>0)$, then the pedal loss is supervised by the relative vehicle state cost $C(\varphi^{(i)}_{t}, \Tilde{p}^{(i)}_{t})
 - C(\varphi^{(i)}_{t}, p^{(i)}_{t})$. This allows the vehicle to speed up when no other agents are nearby and slow down when getting close to other objects or its target. Otherwise, the pedal acceleration loss is supervised by the reference pedal $\Tilde{p}^{(i)}_{t}$. Note that we use the reference steering angle $\varphi^{(i)}_{t}$ rather than the predicted steering angle $\Tilde{\varphi}^{(i)}_{t}$ to calculate the vehicle state cost. 

\subsection{Training \& Test Data Generation}

The advantage of our self-supervised approach is that the data for training the GNN model can be generated on the fly, since it does not require supervised labels. However, for the purpose of reproducibilty of our experiments, we generate data beforehand.  One common option is to generate the samples by choosing a vehicle's starting and target positions randomly on the navigation grid. However, the risk with this approach is that the dataset might be heavily skewed in favour of one scenario and may not capture other types of situations that the vehicle is expected to encounter. In contrast, our training regime is developed to handle these diverse situations. These situations can primarily be segregated into 3 modes: \textit{collision, parking and normal driving}. In the collision mode, the vehicles are placed such that they have a high probability of collision. This is done by first randomly choosing a point on the grid, defined as a "collision center". The starting and target position of at least two vehicles are placed on opposite sides of this collision center with slight random deviation. Hence,
the model will be pushed to learn a collision avoidance maneuver.  Meanwhile, in the parking mode, the target position for each vehicle is sampled near its start position (within a distance of $10$ m). Note that in both the collision and parking modes, the position of the vehicles are also chosen randomly but with certain constraints i.e. starting and target positions being on the opposite sides of collision center (collision mode) or in close proximity (parking mode). In the \textit{normal driving} mode, the starting and target positions are chosen randomly without any of the constraints described earlier. Lastly, for all modes the following additional conditions are to be fulfilled: No two target or two staring positions of vehicles can overlap. Likewise, if there are static obstacles, the starting/target positions cannot overlap with it. The starting position can be placed within an obstacle's circle of influence but not a target position, since then the vehicle will struggle attaining equilibrium. The target position will attract the ego-vehicle whereas the obstacle influence will repel it.  Our training dataset contains training cases from 1 vehicle 0 obstacles to 5 vehicles 8 obstacles, each with 3000 samples that in turn contains 1000 samples from each of the 3 modes.  Each of these 3000 scenarios serve as the starting point for the simulation that is run for $T=200$ timesteps during training, in order for the vehicles to reach their target positions. As mentioned in Section \ref{subsec:model_evaluation}, we also generate test scenarios range from 10 vehicles - 0 obstacles to 50 vehicles - 25 obstacles, each with 1000 samples. However, our test dataset is generated only using the collision mode.

Multiple training samples can be grouped as a batch and run simultaneously. However, this training simulation will always end with all the vehicles in the training batch approaching their targets and parking. To prevent the model from overfitting on parking behavior, we adopt a asynchronous simulation during the training. Specifically, a training batch if further divided into multiple mini-batches. These mini-batches are in different simulation time steps. In this case, within a training batch, there are always some training samples just starting, some driving the vehicles on the way and the rest parking the vehicles. Besides, for each time step $t$ during the training simulation, we also perturbs the vehicle state using a zero-mean Gaussian noise as a data augmentation. The standard variance decreases linearly along the simulation time $t$: $\boldsymbol{\sigma}_{t} = \frac{T-t}{T} \cdot [\sigma_{x}, \sigma_{y}, \sigma_{\varphi}, \sigma_{v}]^{T}$. The validation dataset reuse the training samples without random perturbation. The simulation length during validation is reduced to $T=1$.

To train the neural network, we use the Adam optimizer with an initial learning  rate of $0.01$ and weight decay of $10^{-6}$. The learning rate is reduced by factor of $0.2$ from its previous value, if the validation loss does not reduce for 15 epochs. The number of training epochs is set to be $500$ but training is prematurely stopped, if there is no decrease in the validation loss for $50$ epochs.

\begin{figure*}[ht]
\centering
    \subfigure[\centering{safe rate with no obstacles}]{{\includegraphics[width=0.222\textwidth]{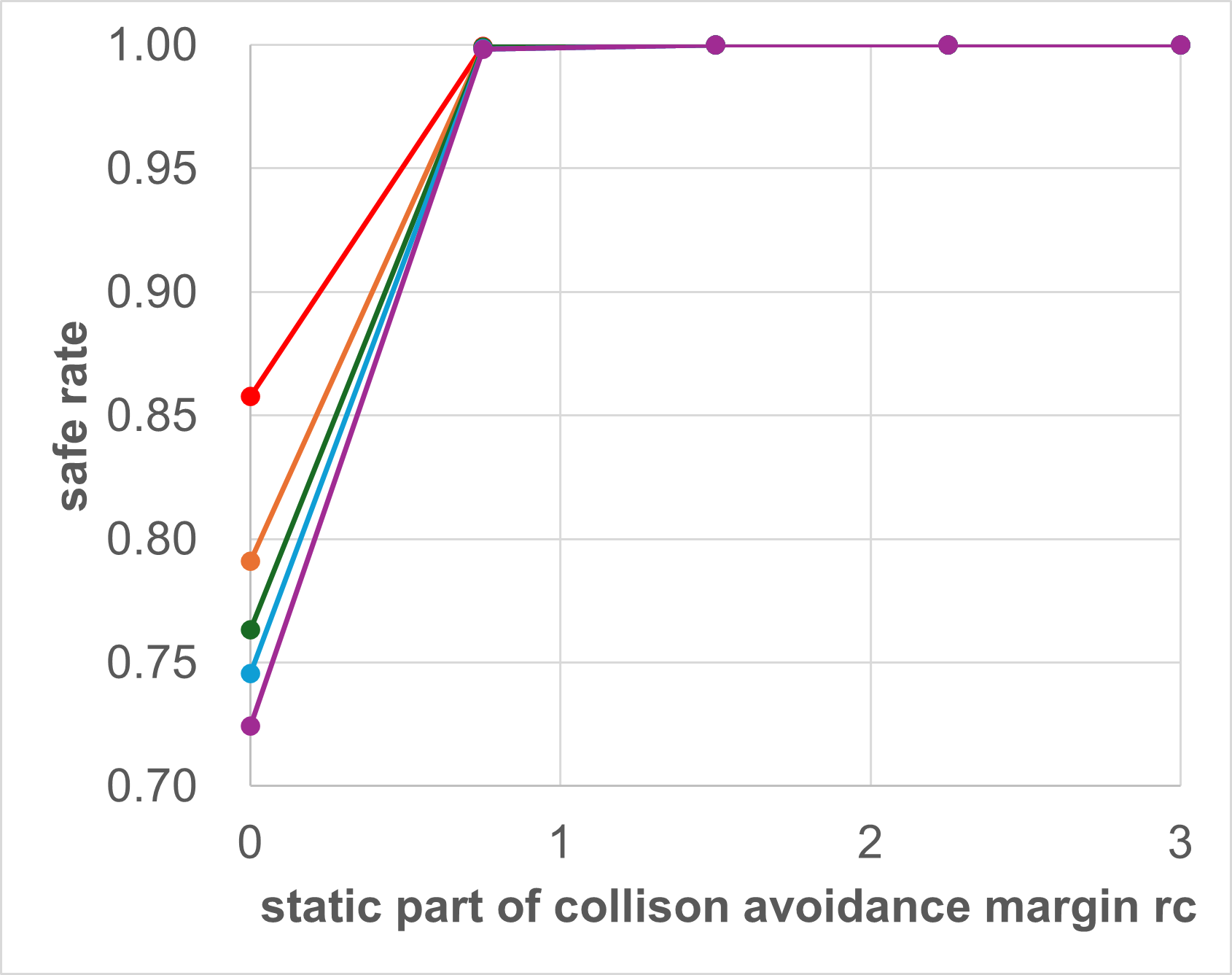}}}
    \subfigure[\centering{safe rate with 25 obstacles}]{{\includegraphics[width=0.222\textwidth]{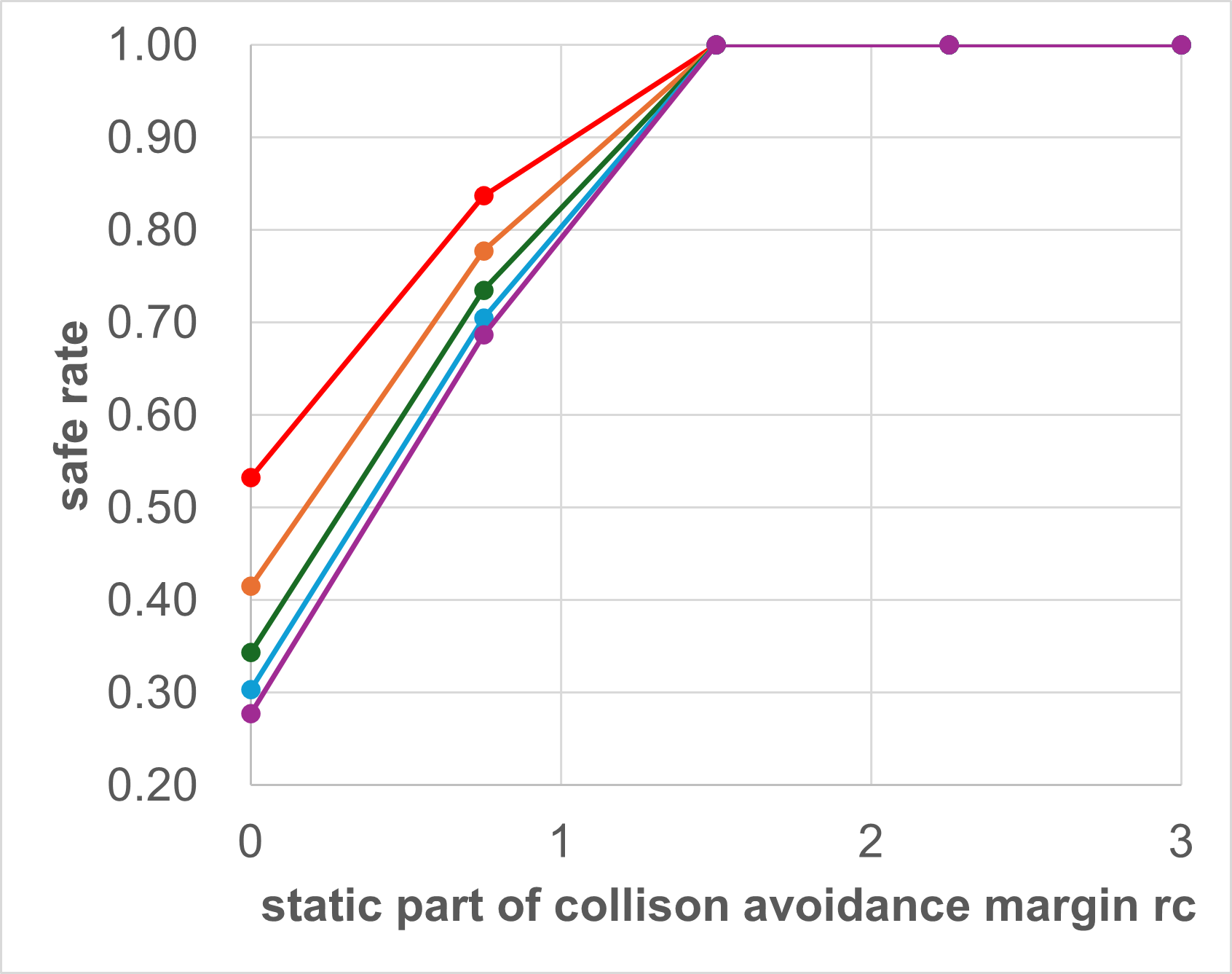}}}
    \subfigure[\centering{reach rate with no obstacles}]{{\includegraphics[width=0.222\textwidth]{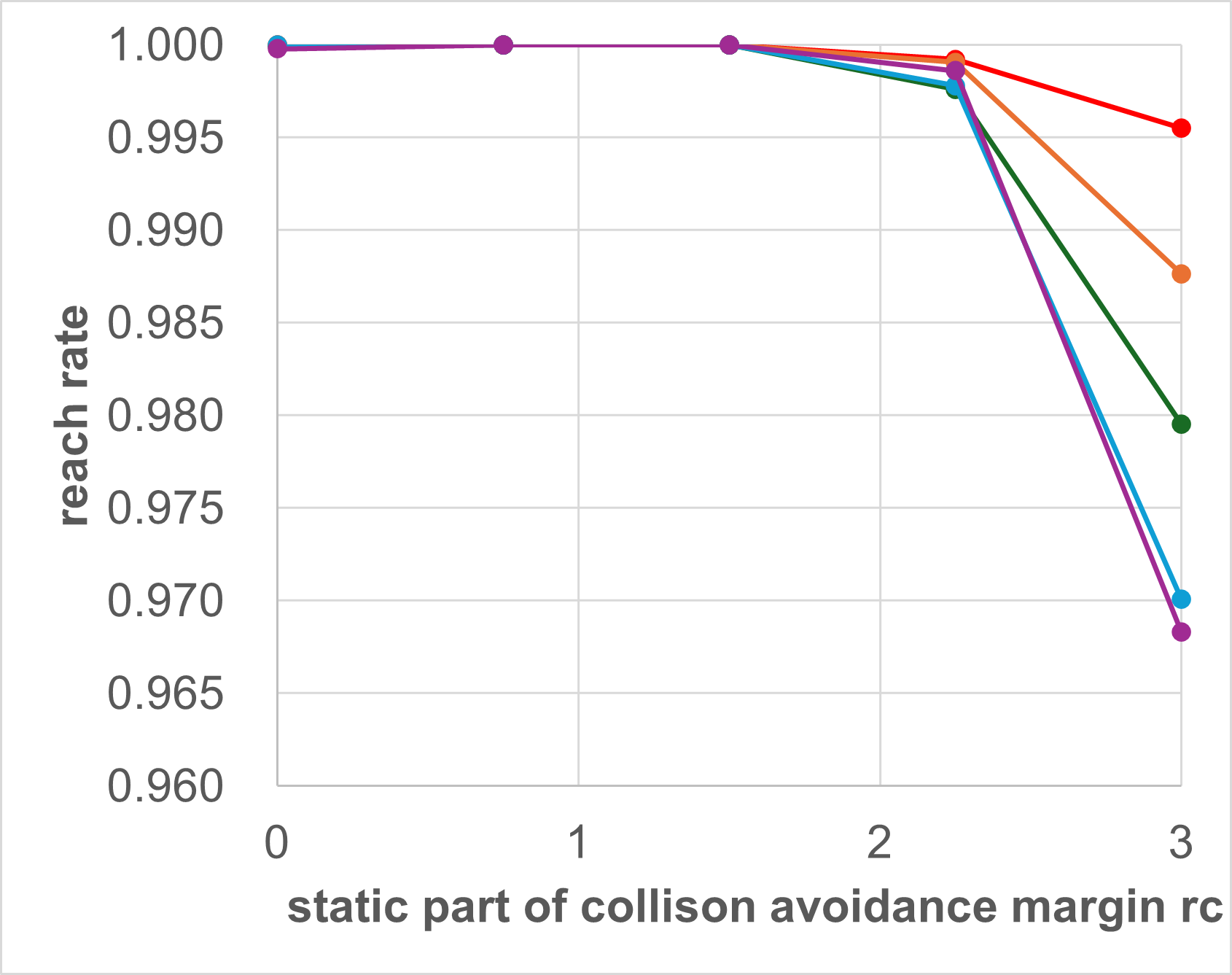}}}
    \subfigure[\centering{reach rate with 25 obstacles}]{{\includegraphics[width=0.222\textwidth]{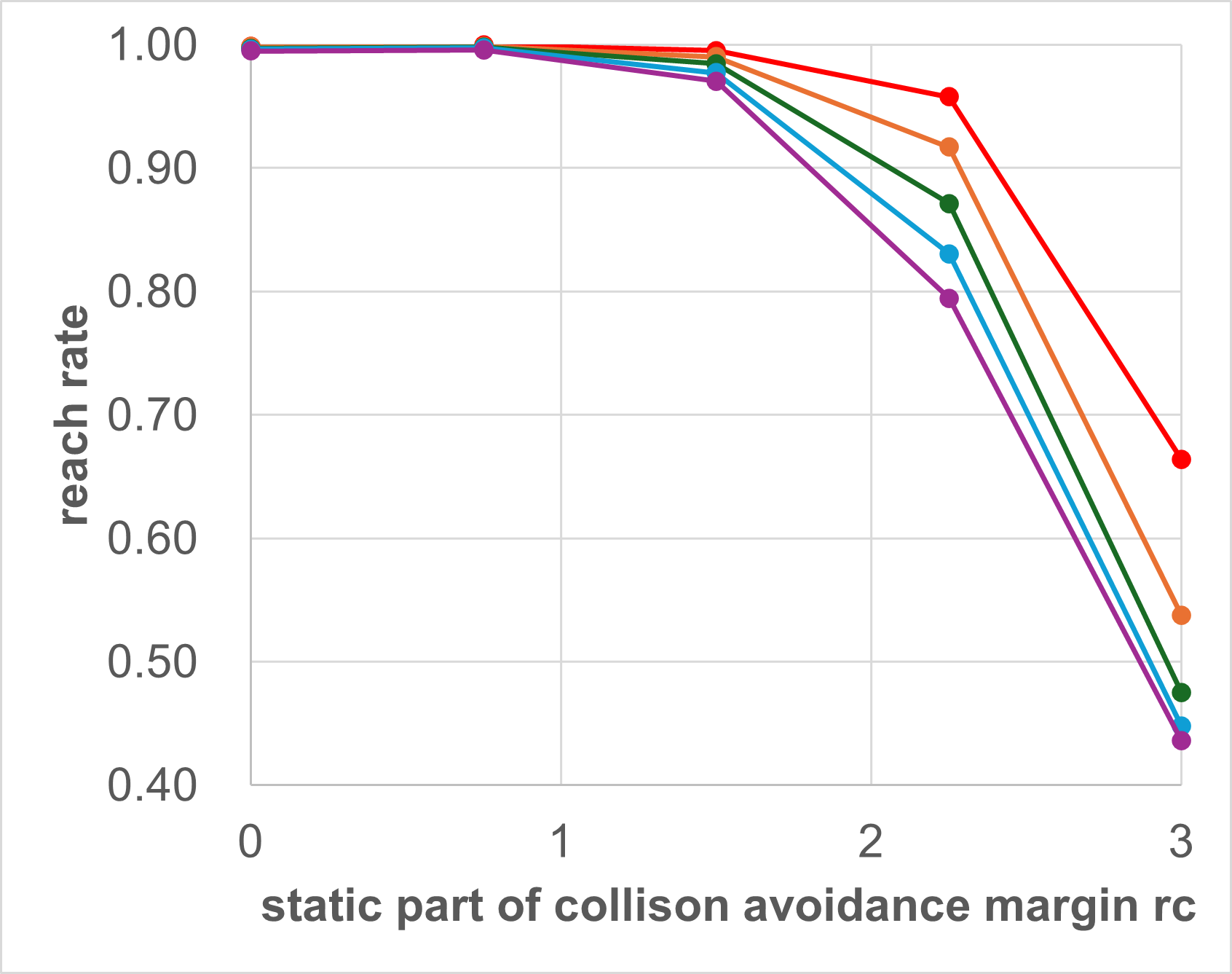}}}
    \subfigure[\centering{Legend}]{{\includegraphics[width=0.078\textwidth]{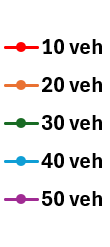}}}
    \caption{Shows the \textit{Safe} and \textit{Reach} rate metrics (Higer is better) for the different static part $r_{c}$ of collision avoidance margin.}
    \label{fig:safe success rate of different r_c}
\end{figure*}

\subsection{Experimental Settings}
In this section, we elaborate the detailed hyper-parameter settings in our experiments.

In the vehicle kinematic Equation \ref{equ:vehicle kinematic}, the inverse of the vehicle length $\gamma$ is $0.5$ and the friction coefficient $\beta$ is $0.99$. The pedal acceleration threshold $P$ introduced in Section \ref{subsec:prob_def} is $1$ m/$s^2$. The steering angle limits  $\Phi$ is $0.8$ radians. The timestep $\Delta t$ is $0.2$ s. 

For the setting of velocity vector field in Subsection \ref{subsec:orientation} and \ref{subsec:speed}, the default velocity $v_{d}$ is $2.5$ m/s. The parking threshold $r_{p}$ is $5$ m. The radius of the obstacle $r^{(k)}_{obs}$ is sampled from $1$ m to $3$ m. The radius of the vehicle $r^{(j)}_{veh}$ is fixed to be $1.5$ m. The $|v^{(i)}_{t}|$ and $|v^{(j)}_{t}|$ in the dynamic part of collision avoidance margin are determined to be the corresponding vehicle speed and the vehicle speed will not exceed the default velocity $v_{d}=2.5$ m/s in our MA-DV$^{2}$F algorithm. The static part of collision avoidance margin $r_{c}$ is fixed to be $1.5$ m. The position tolerance $\epsilon_{p}$ and the orientation tolerance $\epsilon_{o}$ are $0.25$ m and $0.2$ m, respectively. The sensing radius of each vehicle should be large than the threshold determined in Equation \ref{equ:edge filtering threshold}.

\subsection{Sensitivity Analysis}

In this Subsection, we investigate how the collision avoidance margin influences the success rate. As the dynamic part of collision avoidance margin is determined by vehicle speed, we only change the static part $r_{c}$ in our experiment. As can be seen in Table \ref{table:sensitivity analysis success rate} and Fig. \ref{fig:safe success rate of different r_c}, that if the safety margin rc is reduced from its default value of 1.5 m, the success rate reduces drastically. This is because MA-DV2F considers only those neighboring agents (other vehicles and obstacles) that are in its immediate vicinity. Hence, when determining steering commands for the ego-vehicle, MA-DV2F would not consider the presence of those agents that are close but not in the immediate vicinity. Therefore, if the ego-vehicle is moving at high speed, then by the time a neighboring vehicle enters its immediate vicinity, it might have been too late for the ego-vehicle to execute an evasive maneuver away from that neighboring vehicle.  This will lead to higher probability of collision, hence a lower safe rate (Bottom row) and consequently a lower success rate (Top row) as can be seen in the plots in Fig. (a), for all the vehicle configurations (10-50 vehicles).   

 Conversely, if the safety margin rc is increased from its default value of 1.5 m, the success rate also reduces but for a different reason. Note that with a greater safety margin, the number of neighboring agents that would be considered by the MA-DV2F for steering command prediction will also be greater. This will cause the ego-vehicle to behave conservatively in order to prevent collisions at the expense of reaching the target. This results in a lower reach rate (center row) and consequentially a lower success rate (Top row). This is further exacerbated when there are obstacles in the scene (Right column), since the presence of obstacles reduces the region in which the vehicles can navigate without causing collisions.

\begin{table}[!ht]
\centering
\resizebox{\linewidth}{!}{
\begin{tabular}{||c c | c c c c c||} 
\hline
\hline
Number of & Number of & \multicolumn{5}{c||}{static part of collision avoidance margin $r_{c}$} \\
Vehicles & Obstacles & 0.0 & 0.75 &  1.5 & 2.25 & 3.0 \\ [0.5ex]
\hline
\multicolumn{7}{||c||}{success rate $\uparrow$ } \\ 
\hline
10 & 0 & 0.8576 & 0.9994 & \textbf{1.0000} & 0.9992 & 0.9955 \\
\hline
20 & 0 & 0.7910 & 0.9990 & \textbf{1.0000} & 0.9991 & 0.9876 \\
\hline
30 & 0 & 0.7631 & 0.9992 & \textbf{1.0000} & 0.9976 & 0.9795 \\
\hline
40 & 0 & 0.7456 & 0.9984 & \textbf{1.0000} & 0.9978 & 0.9701 \\
\hline
50 & 0 & 0.7244 & 0.9982 & \textbf{1.0000} & 0.9986 & 0.9683 \\
\hline
10 & 25 & 0.5324 & 0.8363 & \textbf{0.9952} & 0.9577 & 0.6637 \\
\hline
20 & 25 & 0.4146 & 0.7764 & \textbf{0.9902} & 0.9171 & 0.5375 \\
\hline
30 & 25 & 0.3434 & 0.7339 & \textbf{0.9844} & 0.8713 & 0.4750 \\
\hline
40 & 25 & 0.3032 & 0.7032 & \textbf{0.9772} & 0.8306 & 0.4480 \\
\hline
50 & 25 & 0.2775 & 0.6843 & \textbf{0.9704} & 0.7945 & 0.4363 \\
\hline
\end{tabular}
}
\caption{Shows how the static part of collision avoidance margin $r_{c}$ effects success rate. }
\label{table:sensitivity analysis success rate}
\end{table}

\end{document}